\documentclass[a4paper,11pt]{article}
\pdfoutput=1
cm \voffset = -2truecm

\usepackage{graphicx}
\usepackage{amsmath}
\usepackage{amssymb}
\usepackage{latexsym}
\usepackage[colorlinks]{hyperref}
\usepackage{color}
\usepackage{cite}
\usepackage{makeidx}
\usepackage{float}

\restylefloat{figure}

\newcommand{\bra}{\begin{array}}
\newcommand{\era}{\end{array}}
\newcommand{\beq}{\begin{equation}}
\newcommand{\eeq}{\end{equation}}
\newcommand{\bqr}{\begin{eqnarray}}
\newcommand{\eqr}{\end{eqnarray}}

\def\BC{\bb C}
\def\_\BC{\bbi C}

\def\no2 {{\textstyle{n\over 2}}}

\newcommand{\lb}{\label}

\begin{document}
\begin{titlepage}
\setcounter{page}{1}
\renewcommand{\thefootnote}{\fnsymbol{footnote}}

\begin{flushright}
\end{flushright}

\vspace{5mm}
\begin{center}

{\Large \bf {Electrostatic and Magnetic Fields in Bilayer Graphene}}

\vspace{5mm}

{\bf Ahmed Jellal\footnote{\sf ajellal@ictp.it --
a.jellal@ucd.ac.ma}}$^{a,b}$, {\bf Ilham Redouani}$^{b}$, {\bf Hocine Bahlouli}$^{a,c}$

\vspace{5mm}

{$^a$\em Saudi Center for Theoretical Physics, Dhahran, Saudi Arabia}


{$^{b}$\em Theoretical Physics Group,  
Faculty of Sciences, Choua\"ib Doukkali University},\\
{\em PO Box 20, 24000 El Jadida, Morocco}

{$^c$\em Physics Department, King Fahd University of Petroleum and Minerals,\\
Dhahran 31261, Saudi Arabia}


\vspace{3cm}

\begin{abstract}

We compute the transmission probability through rectangular
potential barriers and p-n junctions in the presence of a magnetic
and electric fields in bilayer graphene
taking into account contributions from the full four bands of the energy spectrum.
For energy $E$ higher than the interlayer coupling $\gamma_1$ ($E > \gamma_1$) two
propagation modes are available for transport giving rise to four possible
ways for transmission and reflection coefficients. However,
when the energy is less than the height of the barrier the
Dirac fermions exhibit transmission resonances and only one mode
of propagation is available for transport. We study the effect of the interlayer
electrostatic potential denoted by $\delta$ and variations of different barrier geometry parameters
on the transmission probability.

\end{abstract}
\end{center}

\vspace{3cm}


\end{titlepage}


\section{Introduction}

Graphene is a one atom thick single layer of carbon
material, which takes the form of a planar honeycomb lattice of
$sp^2$ bonded carbon atoms. It is the first two-dimensional (2D)
crystalline material which has been experimentally realized
\cite{novoselov2004}. This new material has attractive electronic
properties, among them, an unusual quantum Hall effect
\cite{novoselov2005,Zhang2005} and optical transparency
\cite{Nair2008}. The equation describing the electronic
excitations in graphene is formally similar to the Dirac
equation for massless fermions which travel at a speed of the
order on $10^6 m/s$ \cite{Semenoff,DiVincenzo}. As a result graphene has a
number of {{attractive physical properties which makes it a good candidate for
several applications}}. {{In fact its conductivity can be modified
over a wide range of values either by chemical doping or through the application of a DC}}
electric field. {{The very high mobility of graphene
\cite{Morozov} makes it very attractive for}}
electronic high speed applications \cite{Lin}.

{{Bilayer graphene consists of two single layer graphene sheets stacked in A-B stacking
(also known as Bernal stacking \cite{Bernal}), where the A and B atoms in different layers
are on top of each other. While a single layer graphene has two atoms per unit cell a bilayer
graphene has four atoms per unit cell and atoms in different layers interact with each other.
However, the most important interaction between the two layers is represented by a direct overlap integral
between A and B atoms on top of each other, this interaction is denoted by
$\gamma_1$ \cite{Mac2006}, higher order interactions between other atoms in different layers
will have minor effect on the properties of the bilayer system and
hence will be neglected in the present work.
Many of the properties of bilayer graphene are
similar to those of a single layer graphene \cite{Castro,Manes}.
However, while the energy spectrum of a single layer graphene
consists of two cone shaped bands, bilayer graphene possess
four bands and the lowest conduction and highest valence
bands exhibit quadratic spectra and are tangent to each other near the
K-points \cite{Mac2006,Guinea,Latil,Partoens}. One of the most important applications of bilayer
graphene is the fact that we can easily create and control the energy gap using a static
electric field.}}

Recently there have been some theoretical
investigations on bilayer graphene, in particular the work of Van
Duppen \cite{Duppen} followed our recent work \cite{Hassan}, where we
developed a theoretical model that generalizes \cite{Duppen} and allowed us
to deal with bilayer graphene in the presence of a perpendicular electric and magnetic fields.
A  systematic study revealed that interlayer interaction is
essential, in particular the direct interlayer coupling parameter
$\gamma_1$, for the study of transmission properties. Actually
this interlayer coupling $\gamma_1$ sets the main energy scale in
the problem. For incident energies $E$ we found that for $E <
\gamma_1$ there is only one channel of transmission exhibiting
resonances while for $E > \gamma_1$  two propagating modes are
available for transport resulting in four possible ways of
transmission. Subsequently, we used the transfer matrix method to
determine the transmission probability and associated current
density. This work allowed us to investigate the current density
and transmission through a double barrier system in the presence
of electric and magnetic fields perpendicular to the layers and allowed us to
compare our numerical results with existing literature on the
subject.

The present paper is organized as follows. In
section 2, we formulate our model Hamiltonian
system and compute the associated energy eigenvalues and energy bands.
In section 3, we consider the three potential
regions of the bilayer and obtain the spinor solution
corresponding to each region in terms of barrier parameters and
applied fields. The boundary conditions enabled us to
calculate the transmission and reflection probabiliies. We then
studied two interesting cases corresponding to incident electron energy
either smaller or greater than the interlayer coupling
parameter, $E < \gamma_1$  or $E > \gamma_1$. In section 4 we
consider the first situation where $E < \gamma_1$ which exhibits a two
band tunneling which then results in one transmission and one reflection channel.
Then in section 5 we consider the case $E > \gamma_1$ which leads to a
four band tunneling and results in four transmission and four
reflection channels. In section 6, we show the numerical results
for the conductance and investigate the contribution of each
transmission channel. Finally, in section 7, we conclude our work and summarize
our main results.

\section{Theoretical model}

\begin{figure}[h!]
 \centering
\includegraphics[width=7cm, height=5cm]{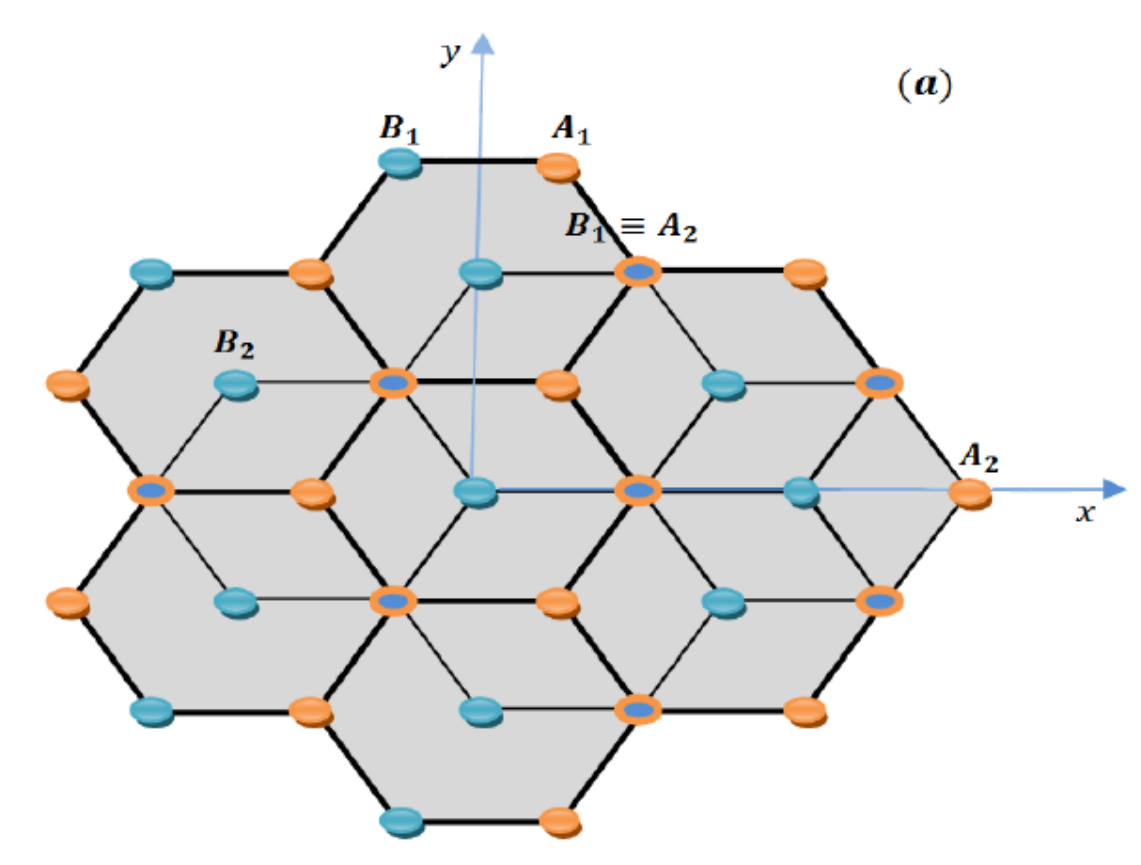}\label{2layer}
\includegraphics[width=8cm, height=6cm]{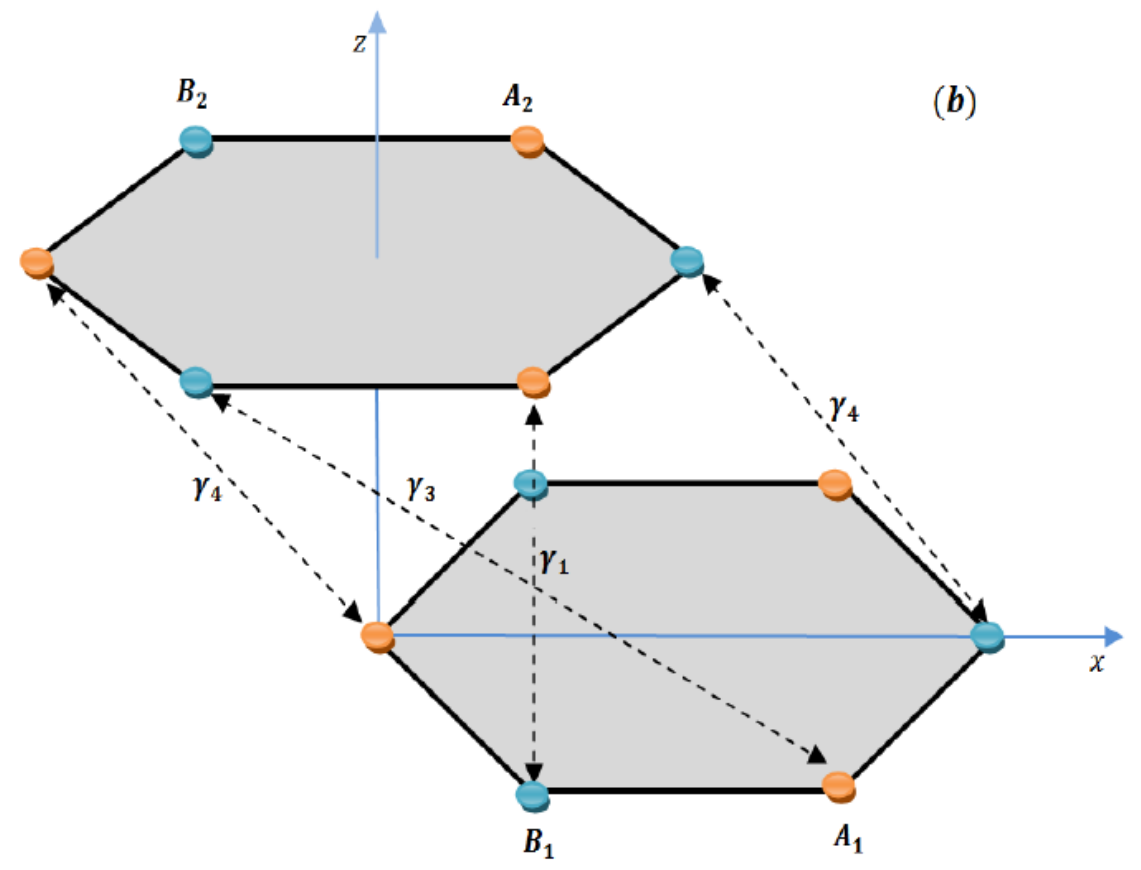}\label{inter}
  \caption{\sf {Lattice structure of bilayer graphene with ($A$,$B$) atoms within the same layer
}}\label{structure}
\end{figure}

{{We consider a bilayer graphene consisting of two A-B stacked layers of
graphene, each layer has two independent basis atoms ($A_{1}$,$B_{1}$) and
($A_{2}$,$B_{2}$), respectively, as shown in Figure
\ref{structure}, where the two indices (1,2) corresponding to the
lower and upper graphene layer, respectively. Every $B_{1}$ site in
the bottom layer lies directly below an $A_{2}$ site in the upper
layer while $A_{1}$ and $B_{2}$ sites do not lie directly below or
above each other. Our theoretical model is based on the well
established tight binding Hamiltonian of graphite \cite{Wallace}
and adopt the Slonczewski-Weiss-McClure parametrization of the
relevant intralayer and interlayer couplings \cite{McClure} to
model our bilayer graphene system. The in-plane hopping parameter,
due to near neighbor overlap, is called $\gamma_0$ and gives rise
to the in-plan carrier velocity. The strongest interlayer
coupling between pairs of $A_2-B_1$ orbitals that lie directly
below and above each other is called $\gamma_1$, this coupling is
at the origin of the high energy bands and plays an important role
in our present work. A much weaker coupling between the $A_1-B_2$
sites, which are not on top of each other, and hence is considered
as a higher order near neighbor interaction leads to an effective
interlayer coupling called $\gamma_3$ the effect of which will be
substantial only at very low energies. The last coupling parameter
$\gamma_4$ represents the interlayer coupling between the same
kind atoms but in different layers $A_1-A_2$ and $B_1-B_2$.
The numerical values of these parameters
have been estimated to be $\gamma_0 \approx 1.4~eV$ for the
intralayer coupling and $\gamma_1 \approx 0.4~eV$ for the most
relevant interlayer coupling while $\gamma_3 \approx 0.3~eV$  and
$\gamma_4 \approx 0.1~eV$.
However, these last two coupling parameters $\gamma_4 $ and $\gamma_3 $
have negligible  effect at high energy and consequently will be neglected
in our present work \cite{Mac2006,Mac2007}.}}

\begin{figure}[h!]
 \centering
\includegraphics[width=12cm, height=4cm]{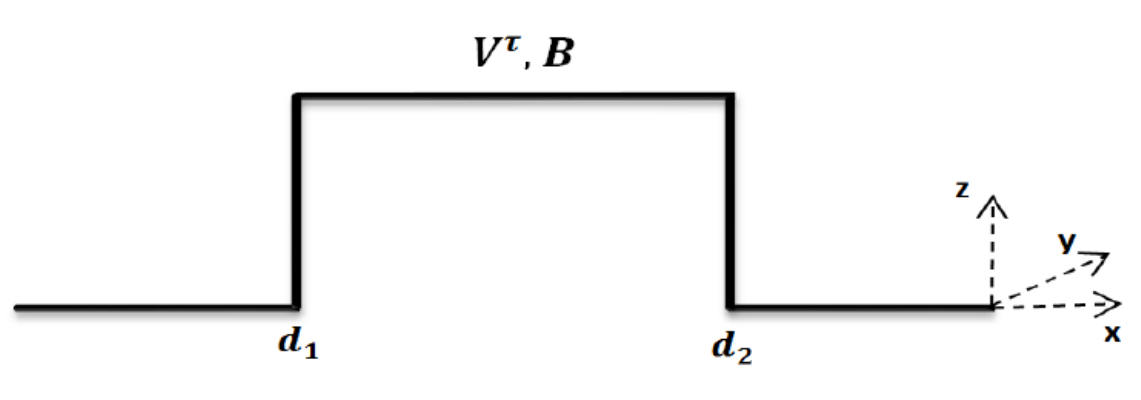}
  \caption{\sf {Schematic diagram for the bilayer graphene barrier.}}\label{structuredeV.}
\end{figure}

\begin{figure}[H]
\centering
\includegraphics[width=6.5cm,
height=5cm]{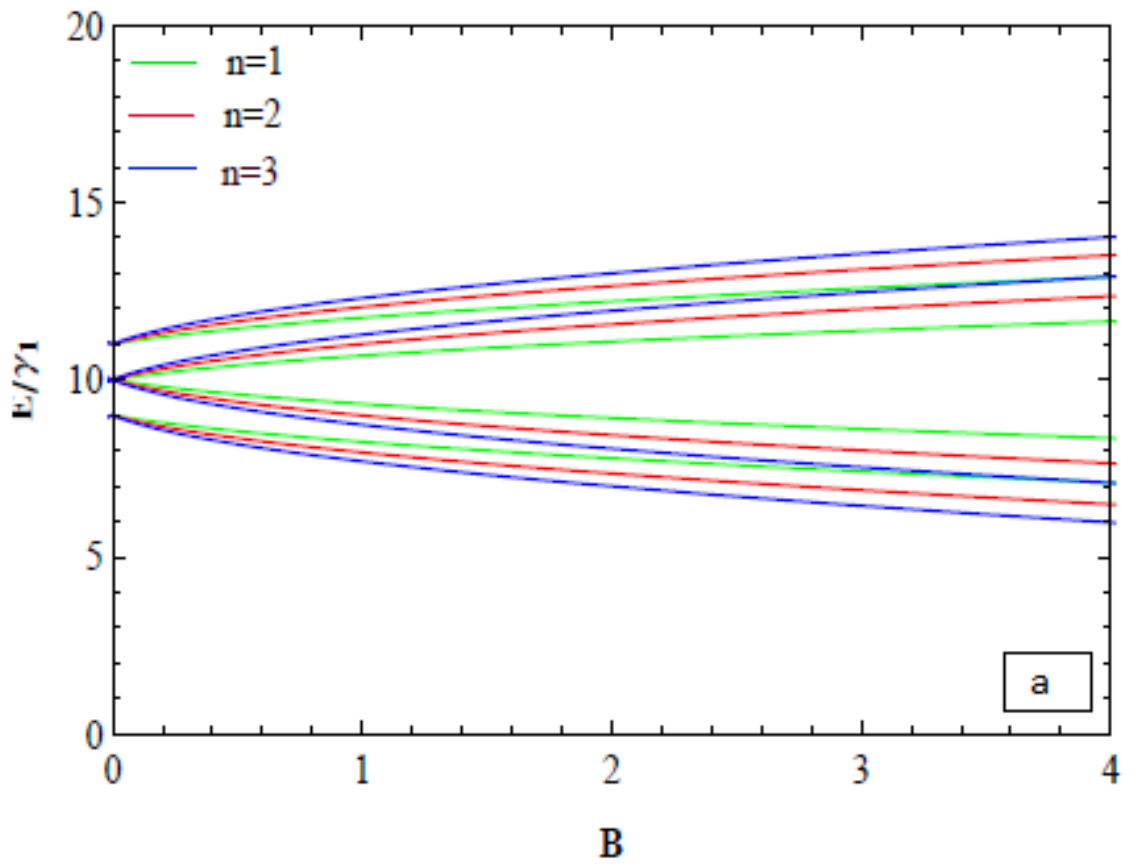}
 \  \
\ \ \ \ \ \ \ \ \ \ \ \
 \includegraphics[width=6.5cm, height=5cm]{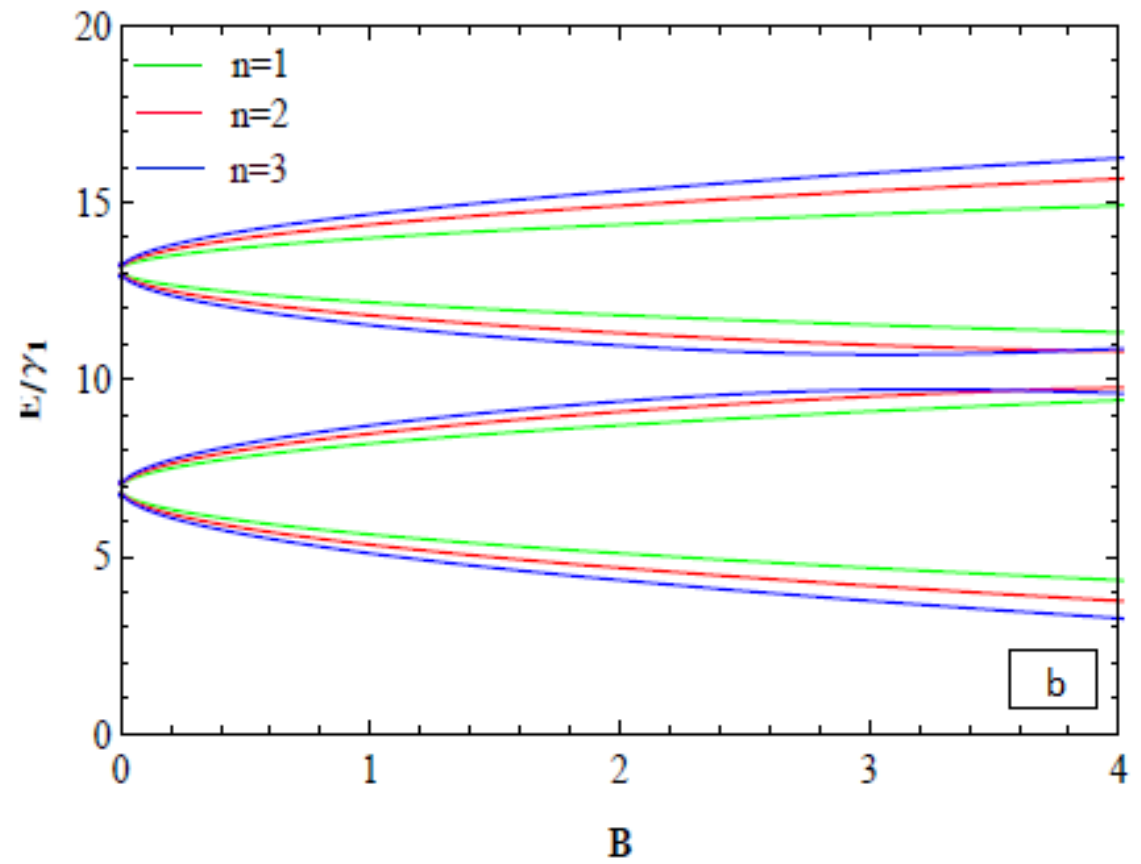}
\caption{The four energy eigenvalues inside the barrier region as
 a function of the magnetic field $B$, with $V=10~\gamma_1$. ({a}) and ({b})
for $\delta=0~\gamma_1$ and $\delta=3~\gamma_1$, respectively.}
\label{eng} \end{figure}

We consider bilayer graphene in the presence of a perpendicular static
electric and  magnetic fields. The charge carriers are scattered by a single
barrier potential along the $x$-direction which results in three different scattering regions
denoted by ${\sf I}, {\sf II}$ and ${\sf III}$. Based on the tight binding approach we can
write the Hamiltonian of the system in the long wavelength limit \cite{Snyman,Neto},
and the associated eigenstates $\psi(x,y)$ as follows
\begin{equation}\label{effective hamiltonien}
H=\left(%
\begin{array}{cccc}
  V^+ & v_F \pi^{+} & -v_4\pi^{+} & v_3\pi \\
  v_F \pi &  V^+  & \gamma_1 & -v_4\pi^{+} \\
  -v_4\pi & \gamma_1 &  V^-  & v_F \pi^{+} \\
   v_3\pi^{+} & -v_4\pi & v_F \pi &  V^-  \\
\end{array}%
\right), \qquad \psi(x,y)=\left(%
\begin{array}{cccc}
  \psi_{A_1}(x,y) \\
   \psi_{B_1}(x,y) \\
   \psi_{A_2}(x,y)  \\
  \psi_{B_2}(x,y) \\
\end{array}%
\right).
\end{equation}
Here $\pi = p_x + i p_y$, $p_{j} = -i \hbar \nabla_{j} + e A_{j}(x,y)$ is the j-th
component of in-plane momentum relative to the Dirac point,
$v_F = \frac{3a}{2}\frac{\gamma_0}{\hbar} =10^{6} m/s$ is the Fermi
velocity for electrons in each graphene layer, $V^+$ and $V^-$ are
the potentials on the first and second layer, and
$v_{3,4}=\frac{v_F\gamma_{3,4}}{\gamma_0}$ are the effective
velocities. We first choose the following potential barrier in
each region as shown in Figure \ref{structuredeV.}, the system is
infinite along the $y$-axis
\begin{equation} \label{eq 2}
V^\tau=\left\{\begin{array}{lll} {0} & \mbox{if} & {x<d_1} \\
{V+\tau\delta} & \mbox{if} & {d_1<x <d_2 } \\
{0} & \mbox{if} & {x>d_2} \end{array}\right.
\end{equation}
{{where $\tau = +1$ for the first layer and $\tau = -1$ for the
second layer so that $2 \delta$ represents the strength of the interlayer
electrostatic potential difference and $V$ is the barrier potential strength.
Choosing the magnetic field to be perpendicular to the
graphene layers, along the $z$-direction and defined by
$B(x,y) = B \Theta \left[(d_1-x)(d_2-x)\right]$ (with constant $B$),
where $\Theta$ is the Heaviside step function. In the Landau
gauge, the corresponding vector potential $A(x,y)=(0,A_y(x))$
giving rise to the above uniform magnetic field takes the form}}
\begin{equation}\label{vector potential}
A_y(x)=\frac{\hbar}{e l_{B}^2}\left\{\begin{array}{lll}
 {d_1} & \mbox{if} & {x<d_1}\\
{x} & \mbox{if} & {d_1<x<d_2}\\
{d_2} & \mbox{if} & {x>d_2}\\
\end{array}\right.
\end{equation}
{{where $l_B = \sqrt{\hbar/e B}$ is the magnetic length and $e$ is
the electronic charge. Since $[H,p_y]=0$ requires conservation of
momentum along the $y$-direction then we can solve the eigenvalue problem
using separation of variables and write the eigenspinors as a plane wave
in the $y$-direction so that our wave function reads}}
\begin{equation}\label{eq2}
\psi(x,y)=e^{ik_yy}\psi(x,k_y)
\end{equation}
At low energies the effect of the parameters v3 and v4 in our original Hamiltonian are negligible on the
transmission coefficient \cite{Duppen}. {Therefore, our Hamiltonian
\eqref{effective hamiltonien} and its associated wavefunction
become}
\begin{equation}\label{eq1}
H=\left(%
\begin{array}{cccc}
  V^+ & v_F \pi^{+} & 0 & 0 \\
  v_F \pi &  V^+  & \gamma_1 & 0 \\
  0 & \gamma_1 & V^-  & v_F \pi^{+} \\
  0 & 0 & v_F \pi &  V^-  \\
\end{array}%
\right), \qquad \psi(x,y)=\left(%
\begin{array}{cccc}
  \psi_{A_1}(x,y) \\
   \psi_{B_1}(x,y) \\
   \psi_{A_2}(x,y)  \\
  \psi_{B_2}(x,y) \\
\end{array}%
\right).
\end{equation}
In the Appendix we solve explicitly our eigenvalue equations and
obtain the following expression for the energy
\begin{eqnarray}\label{e1}
E &=&V+\frac{1}{\sqrt{6}}\left[\pm\left[\mu^{\frac{1}{3}}+(A^2+3C)\mu^{\frac{-1}{3}}+2A\right]^\frac{1}{2}\right.\\
&& \left.\pm\left[-6B\sqrt{6}
\left(\mu^{\frac{1}{3}}+(A^2+3C)\mu^{-\frac{1}{3}}+2A\right)^{-\frac{1}{2}}-\left(\mu^{\frac{1}{3}}+(A^2+3C)\mu^{-\frac{1}{3}}-4A\right)\right]^\frac{1}{2}
\right]\nonumber
\end{eqnarray}
where we defined the quantities
\begin{eqnarray}
&& {\mu=-A^3+27B^2+9AC+3\sqrt{3}\left[-\left(A^2+3C\right)^3+\left(-A^3+27B^2+9AC\right)^2\right]^\frac{1}{2}}\\
&& {A=\delta^2+(2n+1)\vartheta_{0}^2+\frac{\gamma_{1}^2}{2}}\\
&& {B=\vartheta_{0}^2\delta}\\
&& {C=\left((2n+1)\vartheta_{0}^2-\delta^2\right)^2-
\vartheta_{0}^4+\gamma_{1}^2\delta^2}
 \end{eqnarray}
with $\vartheta_0 = \frac{\hbar v_F}{l_B}$ is the energy scale and $n$ is an integer number.
To exhibit the main features of our four energy bands \eqref{e1}, we plot the energy in terms of the  magnetic
field $B$ in Figure \ref{eng}. For $\delta = 0$ and the Landau levels $(n=1,2,3)$,
we observe in Figure \ref{eng}({\color{red}a}) that for {the first and second layers}
we have $E= V$ and $E= V\pm \gamma$, respectively, which correspond to $B=0$.
The situation changes in Figure \ref{eng}({\color{red}b}) when we consider
$\delta \neq 0$ the energy then becomes $E = V\pm \delta$ for $B=0$ and therefore $\Delta
E = 2 \delta $ represents the gap in the energy spectrum. While in both cases, the energy increases/decreases
as long as $B$ and the Landau levels increases inside the barrier.
\noindent For $\delta = 0$, that
is in the absence of electric field,  \eqref{e1} reduces
to \cite{Ramezani}
\begin{equation}
E=V\pm\sqrt{(2n+1)\vartheta_{0}^2+ \frac{\gamma_{1}^2}{2}\pm
\sqrt{(2n+1)\vartheta_{0}^2\gamma_{1}^2+\vartheta_{0}^4+\frac{\gamma_{1}^4}{4}}}.
\end{equation}
These energy eigenvalues will reduce to the case of a single graphene layer where
$\gamma_1 \longrightarrow 0$, to give $E = V \pm \vartheta_{0} \sqrt{2n+1 \pm 1}$.

\begin{figure}[H]
\centering
 \includegraphics[width=6.5cm, height=5cm]{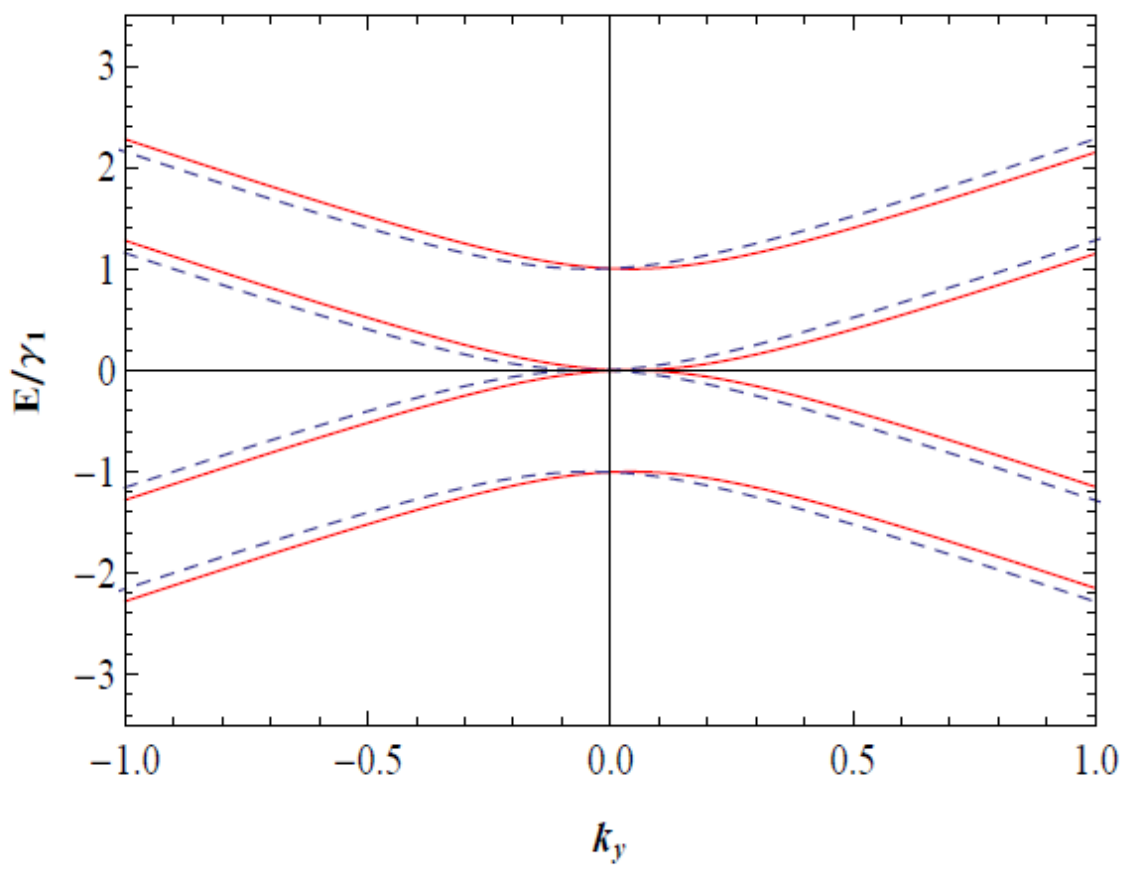}
\caption{The four energy eigenvalues outside the barrier region
as a function of the wave vector $k_y$ along the $y$-direction for
$l_B = 13.5~nm$ and $d_2 = -d_1 = 7.5~nm$, where red (dashed) line
correspond to region ${\sf I}$ (region ${\sf
III}$).} \label{eng1}
\end{figure}

Outside the barrier region, the energy expression can be defined as follows
\begin{equation}\lb{epsl}
\epsilon=
\pm\sqrt{k_{1,2}^2+\frac{\Gamma_{1}^2}{2}\pm\sqrt{\Gamma_{1}^2k_{1,2}^2+\frac{\Gamma_{1}^4}{4}}}
\end{equation}
where $\epsilon = E/ \hbar v_F$, $\Gamma_1 = \gamma_1 / \hbar
v_F$ and
\beq
k_{1,2}=\sqrt{\left(\alpha_{1,2}^\pm\right)^2+\left(k_y+\frac{d_{1,2}}{{l_B}^2}\right)^2}
\eeq
{$\alpha_{1}^\pm$ being} the wave vector of the
propagating wave in the first region where there are two
right-going (incident) propagating modes and two left-going
(reflected) propagating modes. {$\alpha_{2}^\pm$ is} the wave
vector of the propagating wave in the third region with two
right-going (transmission) propagating modes. We plot the energy \eqref{epsl}
in Figure \ref{eng1} to show its behavior in each region
which depends on the propagating modes. It is clear that the behavior is different
in region ${\sf I}$ (red line) and region ${\sf III}$ (dashed line),
as compared to the cases of a simple and double barrier
in the absence of magnetic field \cite{Duppen,Hassan}.\\

Next we will calculate the
transmission and reflection coefficients of electrons across the
potential barrier in our bilayer graphene system.

\section{Transmission probability and conductance}

The transmission and reflection coefficients are obtained by
imposing the continuity of the wave function at each potential
interface. The wave function given in the Appendix can be used in
each region denoted by the integer $j$, {which can then be rewritten in a matrix notation as}
\begin{equation}
\psi_j = G_j \cdot M_j \cdot A_j
\end{equation}
where the index $j$ denotes each potential region, $j = {\sf I}$ for the
incident region ( $ x < d_1 $ ), $j = {\sf II}$ for the potential barrier region
( $ d_1 < x < d_2 $ ) and $j = {\sf III}$ for the transmission region ( $ x > d_2 $ ).
Outside the barrier region, $A_{\sf I}^\pm$ and $A_{\sf III}^\pm$ are defined by
\begin{equation}
 A_{\sf I}^\pm=\left(%
\begin{array}{c}
  \delta_{\pm,1} \\
  r_{+}^\pm \\
  \delta_{\pm,-1} \\
  r_{-}^\pm \\
\end{array}%
\right),\qquad A_{\sf III}^\pm=\left(%
\begin{array}{c}
  t_{+}^\pm \\
  0 \\
  t_{-}^\pm \\
  0 \\
\end{array}%
\right)
\end{equation}
{$\pm$} indicates the wave vector $\alpha_{1,2}^{\pm}$ as defined in the Appendix and $\delta_{\pm,1}$ is the Kronecker
delta function,  $G_{\sf I,III}$ and $M_{\sf I,III}$ are defined by
\begin{equation}
G_{\sf I,III}=\left(%
\begin{array}{cccc}
  f_{1,2}^{++} & f_{1,2}^{+-} & f_{1,2}^{-+} & f_{1,2}^{--} \\
  1 & 1  & 1 & 1 \\
  -1 & -1  & 1 & 1  \\
   -g_{1,2}^{++} & -g_{1,2}^{+-} & g_{1,2}^{-+} & g_{1,2}^{--}   \\
\end{array}%
\right)
\end{equation}
and
\begin{equation}
M_{\sf I,III}=\left(%
\begin{array}{cccc}
  e^{i\alpha_{1,2}^{+}x} & 0 & 0 & 0 \\
  0 &  e^{-i\alpha_{1,2}^{+}x} & 0 & 0 \\
  0 & 0 &  e^{i\alpha_{1,2}^{-}x} & 0 \\
  0 & 0 &  0 &   e^{-i\alpha_{1,2}^{+}x} \\
\end{array}%
\right).
\end{equation}
Inside the barrier region, we have $A_{\sf II}=(c_+,c_-,d_+,d_-)^T$ and
{$G_{\sf II}$} 
\begin{equation}
G_{\sf II}=\left(%
\begin{array}{cccc}
\eta_-\lambda_+\chi_{-1}^{++} &\eta_{-}^*\lambda_+\chi_{-1}^{+-}&
\eta_-\lambda_-\chi_{-1}^{-+}&\eta_{-}^*\lambda_-\chi_{-1}^{--}\\
\chi_{0}^{++} &\chi_{0}^{+-}& \chi_{0}^{-+} &\chi_{0}^{--} \\
\zeta^+\chi_{0}^{++} &\zeta^+\chi_{0}^{+-}& \zeta^-\chi_{0}^{-+} &\zeta^-\chi_{0}^{--} \\
\eta_{+}^*\zeta^+\chi_{1}^{++} &\eta_{+}\zeta^+\chi_{1}^{+-}&
\eta_{+}^*\zeta^-\chi_{1}^{-+} &\eta_{+}\zeta^-\chi_{1}^{--}
\end{array}
\right)
\end{equation}
where $\chi_{l}^{\pm\pm}=D[\lambda_\pm+l,\pm Z]$ and $M_{\sf II}=\mathbb{I}_{4}$. 

{The continuity boundary conditions at $x = d_1$ and
$x = d_2$ can be written in a matrix notation as}
\begin{eqnarray}
&& {G_{\sf I}\cdot M_{\sf I}(x=d_1)\cdot A_{\sf I}^\pm=G_{\sf II}(x=d_1)\cdot M_{\sf II}\cdot
A_{\sf II}}\\
&& {G_{\sf III}\cdot M_{\sf III}(x=d_2)\cdot
A_{\sf III}^\pm=G_{\sf II}(x=d_2)\cdot M_{\sf II}\cdot A_{\sf II}}.
\end{eqnarray}
Using the transfer matrix method we can connect $A_{\sf I}^\pm$ with
$A_{\sf III}^\pm$ through {the matrix $N$}
\begin{equation}
N=M_{\sf I}^{-1}(x=d_1)\cdot G_{\sf I}^{-1}\cdot G_{\sf II}(x=d_1)\cdot
G_{sf II}^{-1}(x=d_2)\cdot G_{\sf III}\cdot M_{\sf III}(x=d_2)
\end{equation}
with the help of the relation $A_{\sf I}^\pm = N A_{\sf III}^\pm$, the transport coefficients {can then be derived from}
\begin{equation}
\left(%
\begin{array}{c}
 t_{+}^\pm \\
  r_{+}^\pm \\
  t_{-}^\pm \\
 r_{-}^\pm \\
\end{array}%
\right)
=\left(%
\begin{array}{cccc}
  N_{11} & 0 & N_{13} & 0 \\
  N_{21} & -1 & N_{23} & 0 \\
  N_{31} & 0 & N_{33} & 0 \\
  N_{41} & 0 & N_{43} & -1 \\
\end{array}%
\right)^{-1}\cdot \left(%
\begin{array}{c}
  \delta_{\pm,1} \\
  0 \\
  \delta_{\pm,-1} \\
  0 \\
\end{array}%
\right)
\end{equation}
where $N_{ij}$ are the matrix elements of the matrix $N$. {Then the transmission and reflection coefficients can be
obtained as}
\begin{eqnarray}
&&{t_{+}^\pm=\frac{N_{13}\delta_{\pm,-1}-N_{33}\delta_{\pm,1}}{N_{13}N_{31}-N_{33}N_{11}}}\\
&& {t_{-}^\pm=\frac{-N_{11}\delta_{\pm,-1}+N_{31}\delta_{\pm,1}}{N_{31}N_{13}-N_{11}N_{33}}}\\
&& {r_{+}^\pm=N_{21}t_{+}^\pm+N_{23}t_{-}^\pm}\\
&& {r_{-}^\pm=N_{41}t_{+}^\pm+N_{43}t_{-}^\pm}.
\end{eqnarray}

{On the other hand, the transmission and
refection probabilities can be obtained using the
current density corresponding to our system. This is}
\begin{equation}\label{currentdensite}
\vec{J}=\pm i \psi^\dagger(x,k_y)
\vec{\sigma}\psi(x,k_y)
\end{equation}
where $J$ defines the electric current density for our system.
Computing explicitly equation \eqref{currentdensite} gives for the
incident, reflected and transmitted current densities
\begin{eqnarray}
&& {J_{x}^{\sf inc}=\pm 4 i
\frac{\alpha_{1}^\pm}{\epsilon}}\\
&& {J_{x}^{\sf ref}=\mp 4
i\frac{\alpha_{1}^\pm}{\epsilon} (r_{\pm}^\pm)^* r_{\pm}^\pm
}\\
&& {J_{x}^{\sf tra}=\pm 4 i\frac{\alpha_{2}^\pm}{\epsilon}
(t_{\pm}^\pm)^* t_{\pm}^\pm}
\end{eqnarray}
{which gives rise to the probabilities}
\begin{eqnarray}
&&
{T_{\pm}^\pm=\frac{\mid{J_{x}^{\sf tra}\mid}}{\mid{J_{x}^{\sf inc}\mid}}=\frac{\alpha_{2}^\pm}{\alpha_{1}^\pm}\mid
 t_{\pm}^\pm\mid^2}\\
 && {R_{\pm}^\pm=\frac{\mid{J_{x}^{\sf ref}\mid}}{\mid{J_{x}^{\sf inc}\mid}}=\frac{\alpha_{1}^\pm}{\alpha_{1}^\pm}\mid
 r_{\pm}^\pm\mid^2}.
\end{eqnarray}
Therefore, we ended up with four transport channels for transmissions and reflections
probabilities because we have four bands.
Since electrons can be scattered into four propagation modes then we need to
take into account the change in their wave velocities.
The conductance of our system can be expressed in terms of
the transmission probability using the famous
Landauer-B$\ddot{u}$ttiker formula \cite{Blanter}
\begin{equation}
G=G_0\frac{
L_y}{2\pi}\int_{-\infty}^{\infty}dk_y\sum_{\pm,\pm}T_{\pm}^\pm\left(E,k_y\right)
\end{equation}
where $G_0=Ne^2/(2\pi\hbar)\approx 3.87\times
10^{-5}N\Omega^{-1}$, $N$ is the number of transverse channels and
$L_y$ is the width of the sample in the $y$-direction.

We will investigate numerically two interesting cases depending on the
value of the incident energies, $E$, as compared with the
interlayer coupling parameter $\gamma_1$. The two band tunneling
leads to one transmission and one reflection channel, {takes
place} at energies less than the interlayer coupling ($E <
\gamma_1$) since we have juste one mode of propagation $\alpha^+$. On
the other hand, for energies higher than the interlayer coupling
parameter $\gamma_1$ ($E > \gamma_1$), the four band tunneling
takes place {and gives} rise to four transmission and four reflection
channels. We denote them as $T_{+}^{+}$ and $T_{-}^{-}$ for
scattering from the $\alpha^+$ and $\alpha^-$, respectively.
Therefore, we have two transmission channels ($T_{-}^{+}$ and
$T_{+}^{-}$) of electrons moving in opposite direction (from
$\alpha^+$ to $\alpha^-$ and $\alpha^-$ to $\alpha^+$). In the
next sections we will study each of these regimes separately.
For numerical convenience we fix $\vartheta_0/\gamma_1 = 1.64/l_B$ in the rest of the paper.

\section{Two Band Tunneling}

\begin{figure}[H]
 \centering
 \includegraphics[width=6.5cm, height=4.5cm]{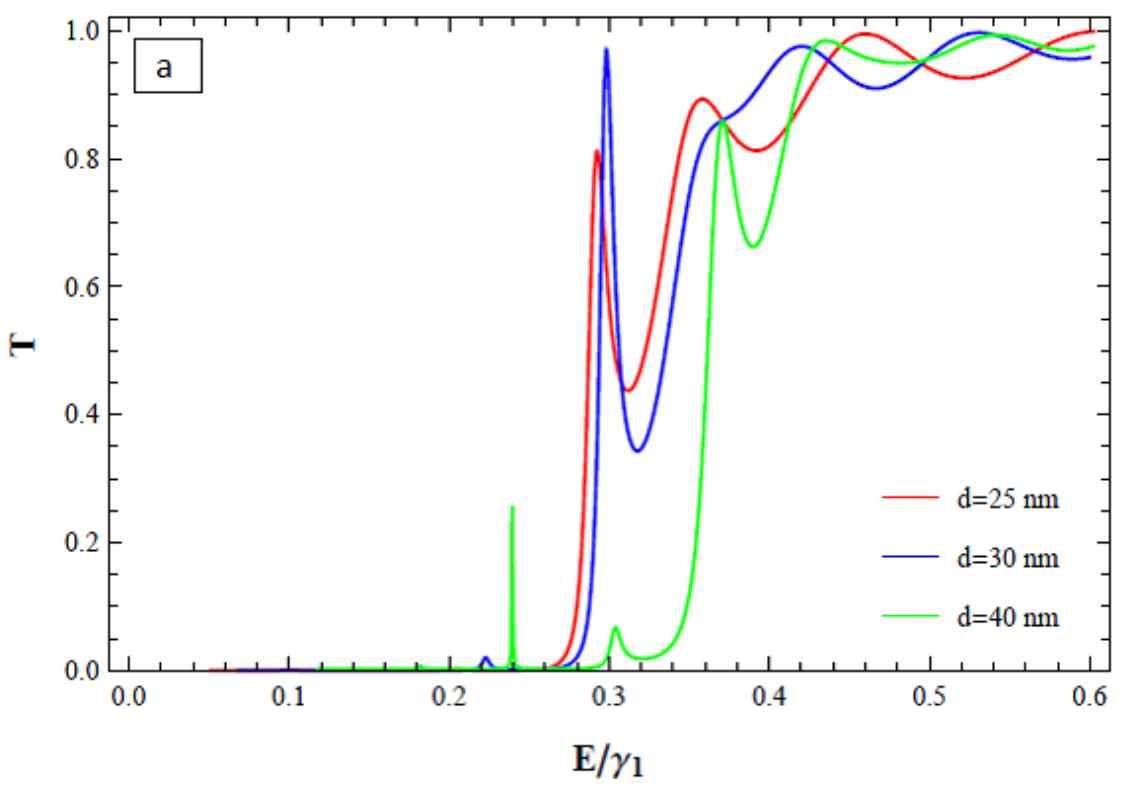}
\ \
 \includegraphics[width=6.5cm,height=4.5cm]{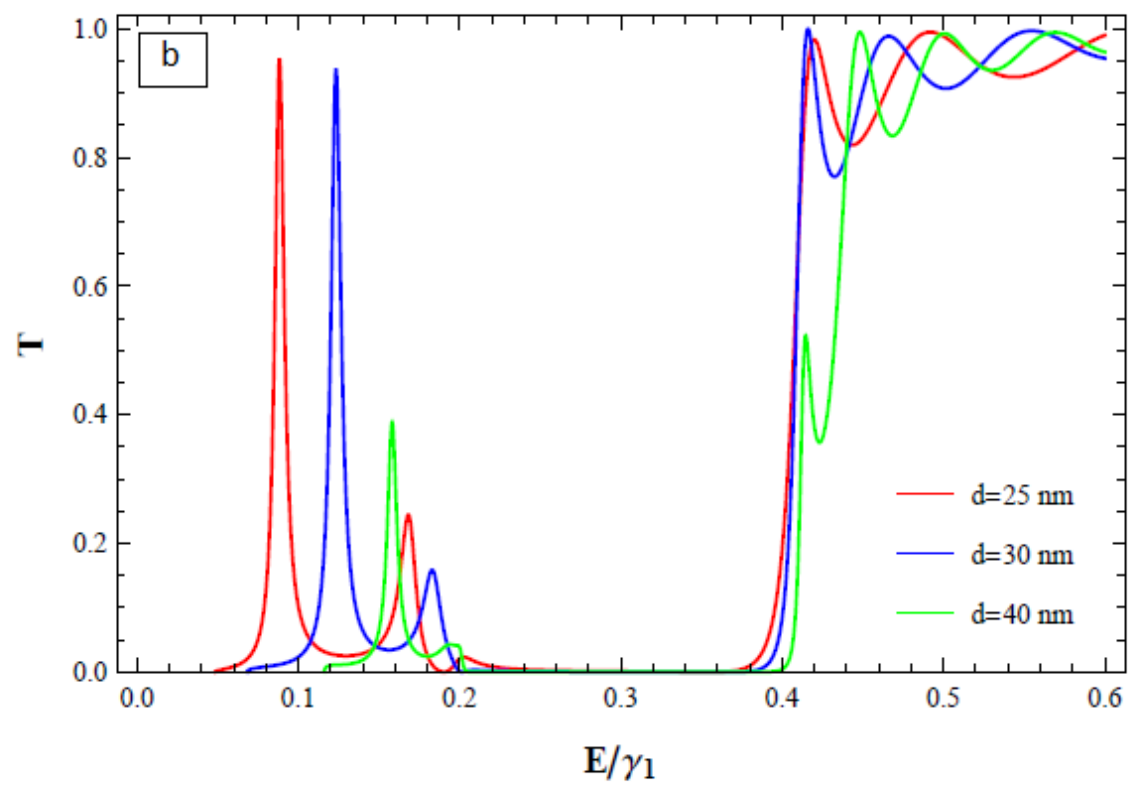}
 \\
 \includegraphics[width=6.5cm, height=4.5cm]{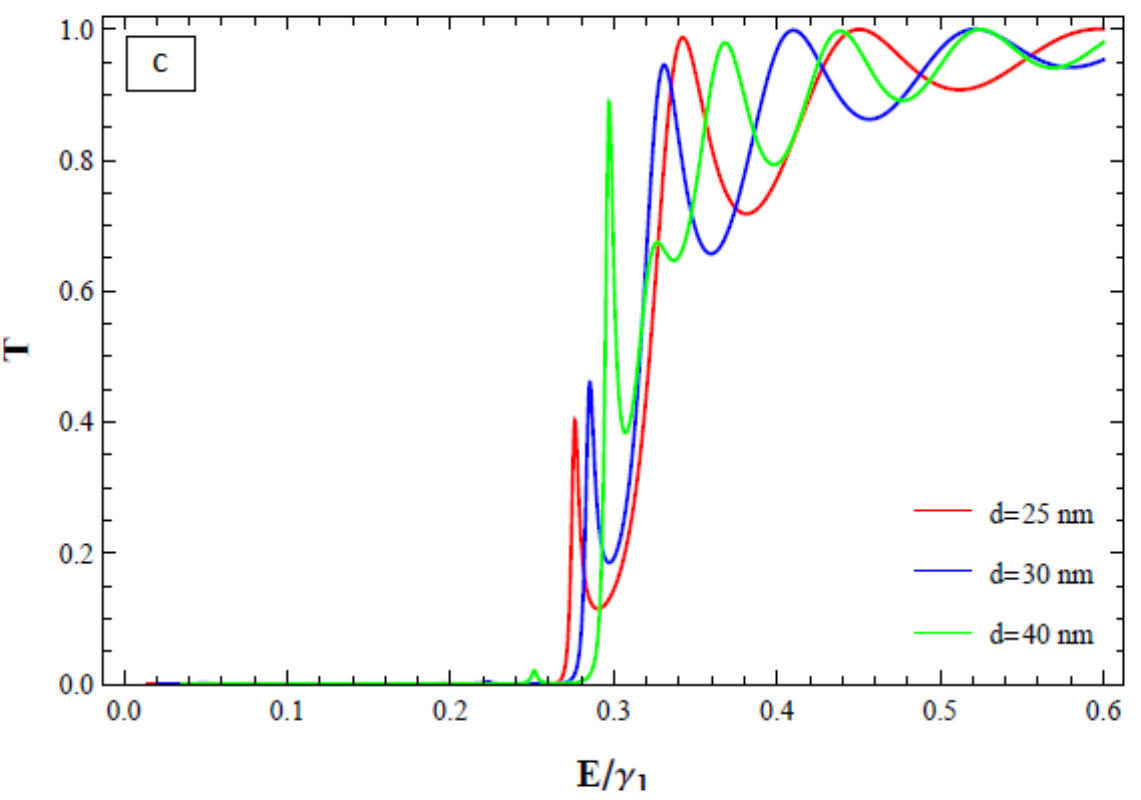}
\ \
 \includegraphics[width=6.5cm, height=4.5cm]{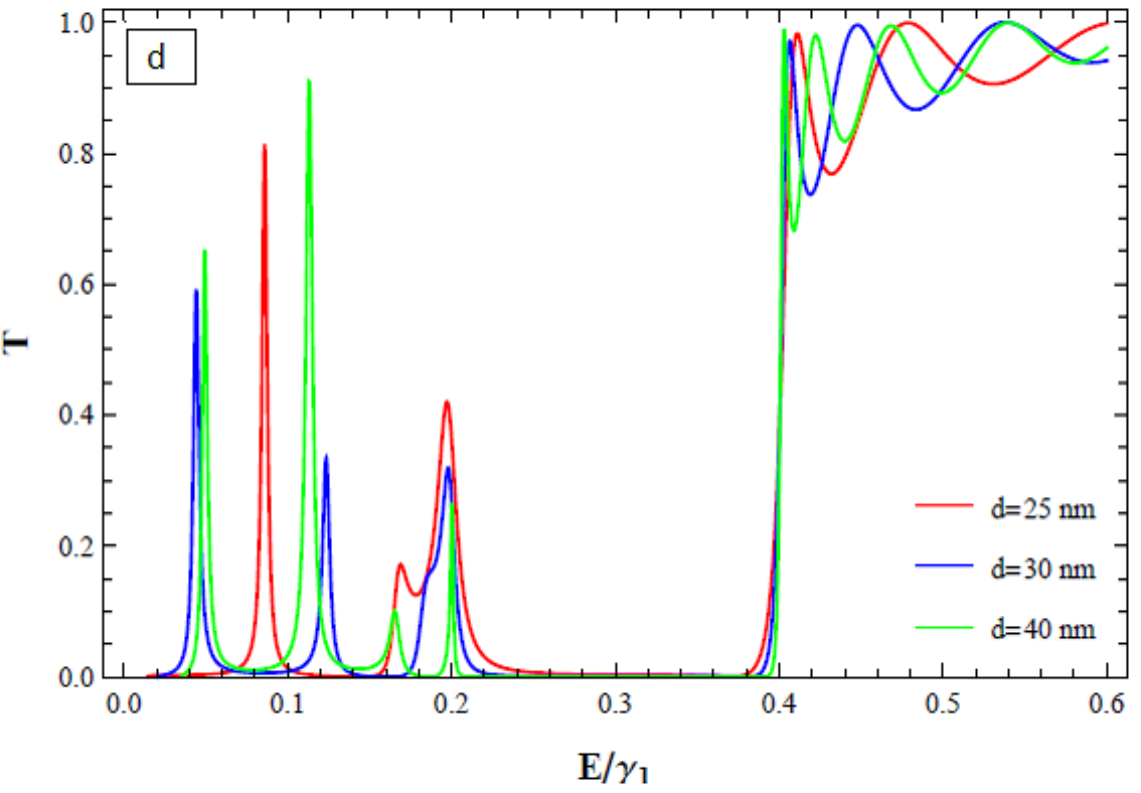}
\caption{Plot of transmission probability as a function of the
Fermi energy $E$ at normal incidence, for $V=0.3~\gamma_1$.
(\textcolor[rgb]{0.98,0.00,0.00}{a})/(\textcolor[rgb]{0.98,0.00,0.00}{b})
for $\delta=0.0~\gamma_1$/$\delta=0.1~\gamma_1$ and $l_B=13.5~nm$
.
(\textcolor[rgb]{0.98,0.00,0.00}{c})/(\textcolor[rgb]{0.98,0.00,0.00}{d})
for $\delta=0.0~\gamma_1$/$\delta=0.1~\gamma_1$ and
$l_B=18.5~nm$.}\label{fig.T2B1}
 \end{figure}

 To allow for a suitable interpretation of our main results in the
low energy regime ($E < \gamma_1$), we compute numerically the
transmission probability under various conditions. First we plot
the transmission probability at normal incidence ($k_yl_B =
-d_1/l_B = 0$) as a function of the Fermi energy $E$, for $V =
0.3~\gamma_1$ {and} three different values of the barrier width
$d=25~nm$ (red line), $d=30~nm$ (blue line), and $d=40~nm$ (green
line), see Figure \ref{fig.T2B1}. {Note that} Figures
(\textcolor[rgb]{0.98,0.00,0.00}{a})/(\textcolor[rgb]{0.98,0.00,0.00}{b})
have been produced for $\delta=0.0~\gamma_1$/$\delta=0.1~\gamma_1$ and $l_B=13.5~nm$
while Figures
(\textcolor[rgb]{0.98,0.00,0.00}{c})/(\textcolor[rgb]{0.98,0.00,0.00}{d})
were done for $\delta=0.0~\gamma_1$/$\delta=0.1~\gamma_1$ and $l_B=18.5~nm$.
We note that in Figure
\ref{fig.T2B1}(\textcolor[rgb]{0.98,0.00,0.00}{a}), when the
energy is less than the height of the barrier potential, i.e $E <
V$, we have zero transmission, while, when the energy is more then
the height of the barrier Dirac fermions exhibit transmission
resonances. As usual the transmission probability  is slightly
displaced to the left as we increase the width of the barrier.
Figure \ref{fig.T2B1}(\textcolor[rgb]{0.98,0.00,0.00}{b}) shows
the transmission for the same parameters
\ref{fig.T2B1}(\textcolor[rgb]{0.98,0.00,0.00}{a}) but with
$\delta = 0.1~\gamma_1$. It is clear that the transmission
probability is affected by the transmission gap $\Delta E = 2
~\delta$. To understand more accurately our system and study the effect of
the magnetic length parameters $l_B$ on the transmission as
function of the Fermi energy $E$ for different values of the magnetic length parameters.
Using the barrier parameters used in Figure
\ref{fig.T2B1}(\textcolor[rgb]{0.98,0.00,0.00}{a}) but with
$l_B=18.5~nm$, see show in Figure
\ref{fig.T2B1}(\textcolor[rgb]{0.98,0.00,0.00}{c}) and
\ref{fig.T2B1}(\textcolor[rgb]{0.98,0.00,0.00}{d}) we show the transmission for zero gap
$\delta=0.0~\gamma_1$ and finite gap $\delta=0.1~\gamma_1$,
respectively. We can clearly see that as we increase $l_B$ the
transmission resonances increase in number while the transmission
probability exhibit a translation to left as we increase the barrier width.\\

\begin{figure}[H]
\centering
\includegraphics[width=5.5cm, height=4cm]{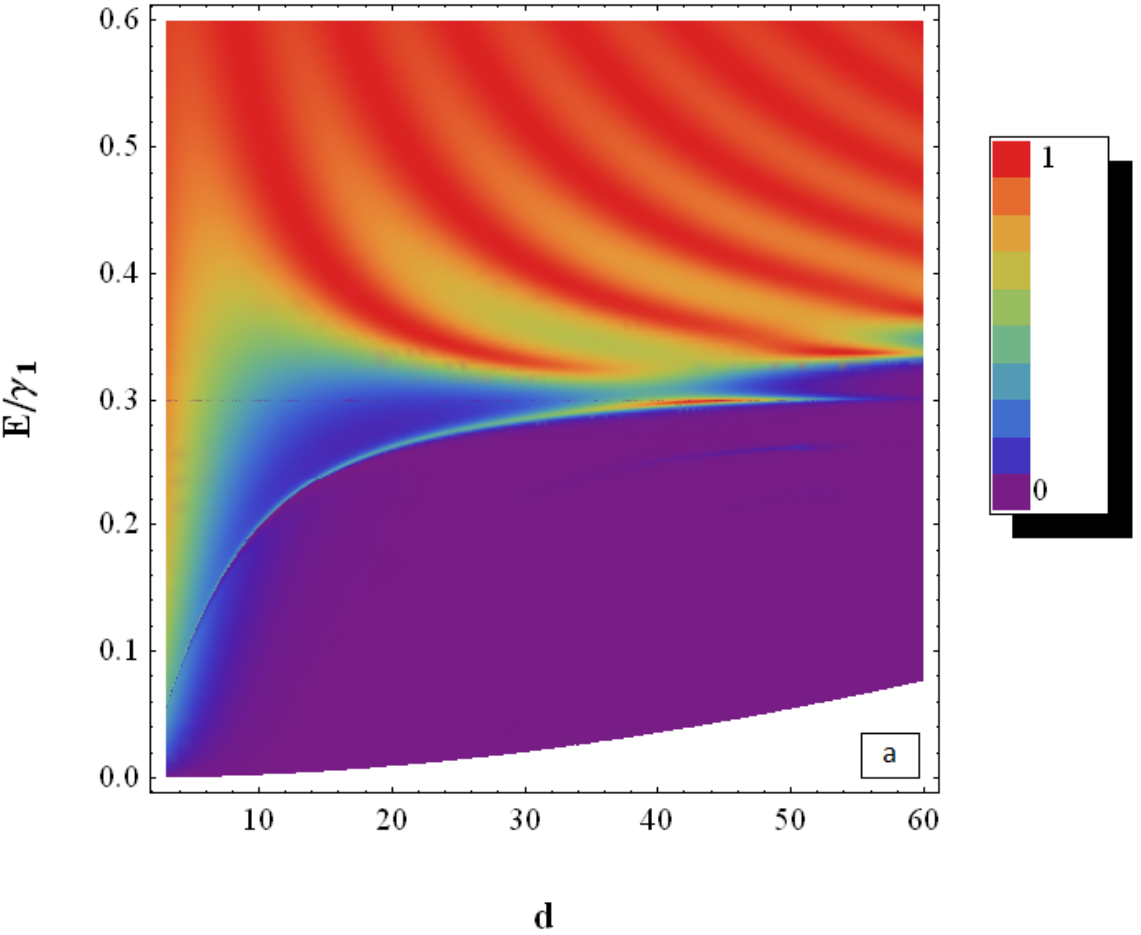}
\ \
\includegraphics[width=5.5cm, height=4cm]{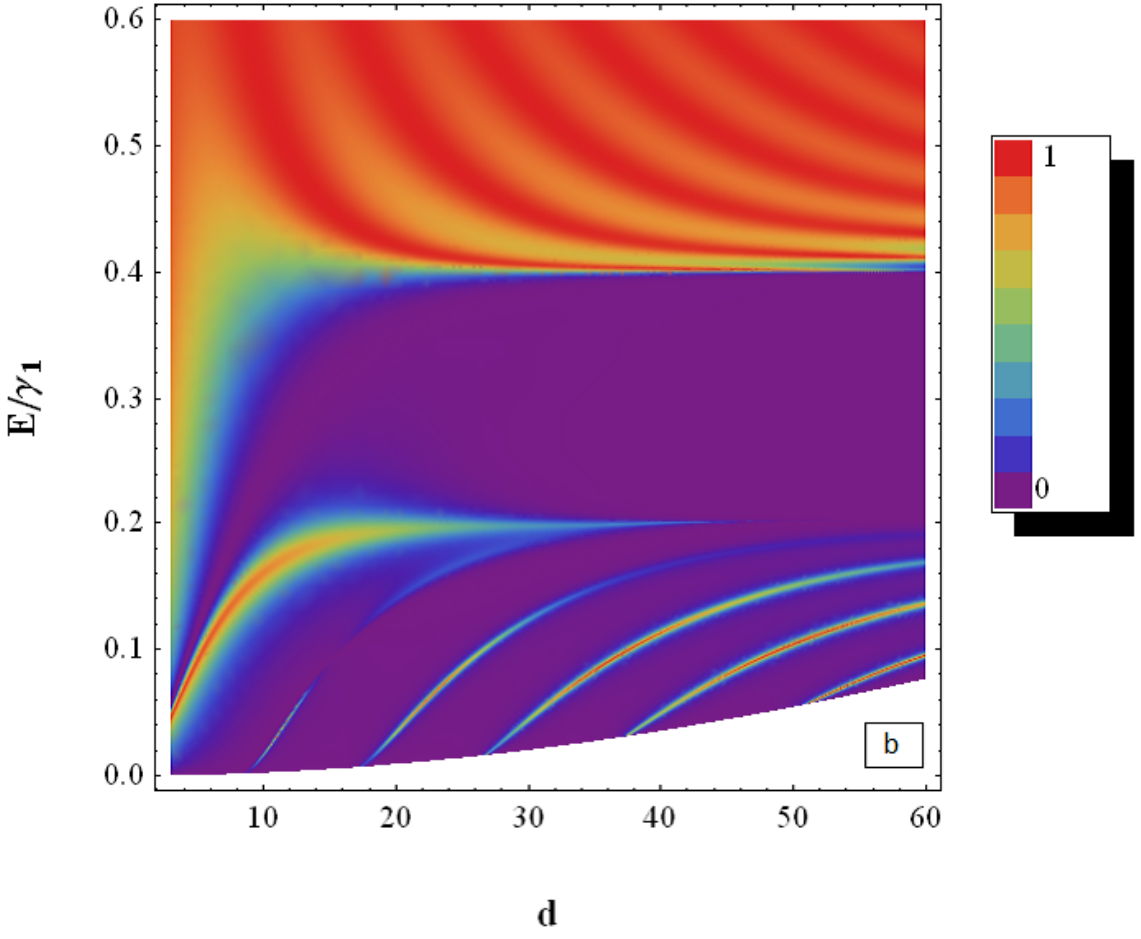}
\\
\includegraphics[width=5.5cm, height=4cm]{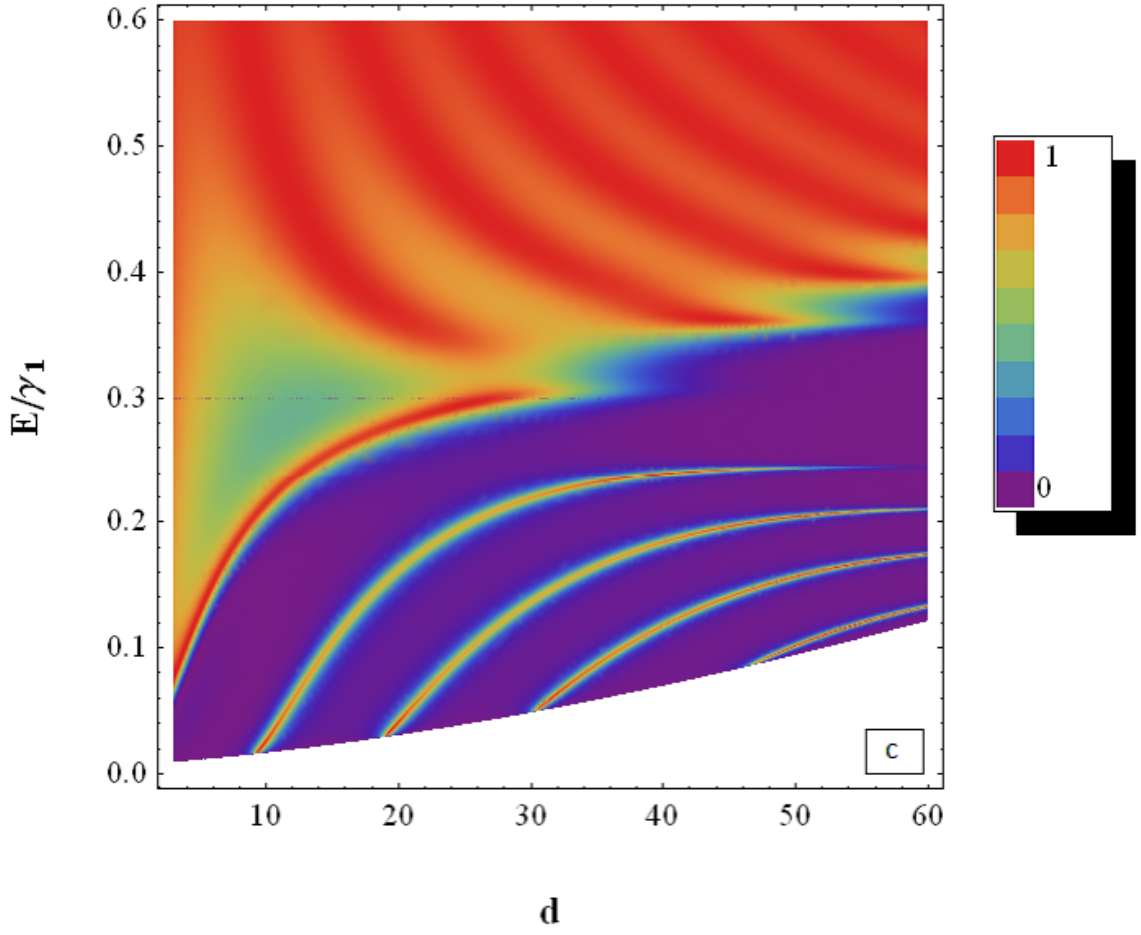}
\ \
\includegraphics[width=5.5cm, height=4cm]{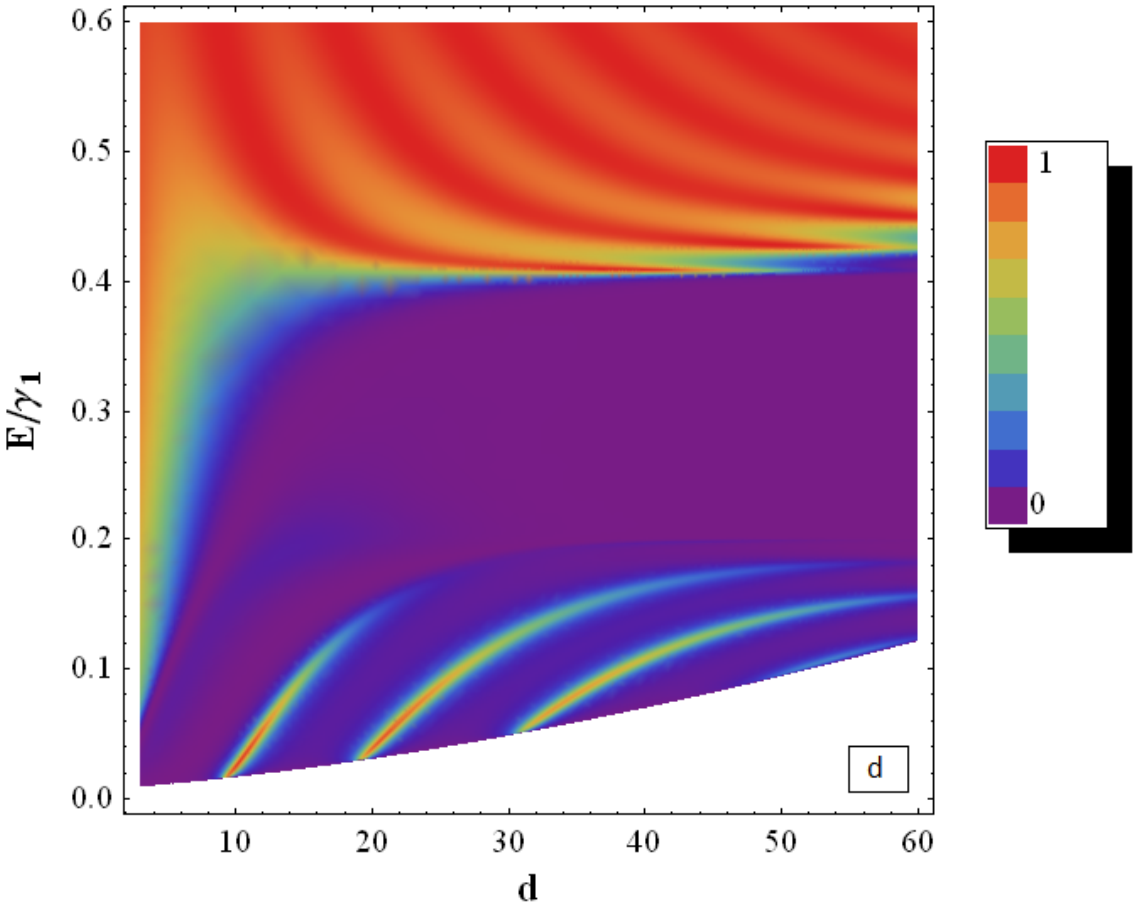}
\caption{Density plot of transmission probability as a function of
the barrier width $d$ and energy $E$, for $V = 0.3~\gamma_1$ and
$l_B=18.5~nm$.
(\textcolor[rgb]{0.98,0.00,0.00}{a})/(\textcolor[rgb]{0.98,0.00,0.00}{b})
for $\delta=0.0~\gamma_1$/$\delta=0.1~\gamma_1$ at normal
incidence $k_y l_B=-d_1/l_B=0$.
(\textcolor[rgb]{0.98,0.00,0.00}{c})/(\textcolor[rgb]{0.98,0.00,0.00}{d})
for $\delta=0.0~\gamma_1$/$\delta=0.1~\gamma_1$ at non-normal
incidence $k_y l_B\neq-d_1/l_B$($d_1=0~nm$ and
$ky=0.05~nm^{-1}$).}\label{fig.6}
\end{figure}

 Figures \ref{fig.6}(\textcolor[rgb]{0.98,0.00,0.00}{a}) and
\ref{fig.6}(\textcolor[rgb]{0.98,0.00,0.00}{c}) show a comparison
of the density plots for the transmission probability at normal
incidence $k_y l_B=-d_1/l_B=0$ and non-normal incidence $k_y l_B
\neq -d_1/l_B$ ($d_1=0~nm$ and $k_y=0.05~nm^{-1}$), as a function
of the barrier width $d$ and energy $E$, respectively, for
$\delta=0.0~\gamma_1$ and $V=0.3~\gamma_1$ in both cases. In
Figures \ref{fig.6}(\textcolor[rgb]{0.98,0.00,0.00}{b}) and
\ref{fig.6}(\textcolor[rgb]{0.98,0.00,0.00}{d}) we used the same
parameters as in \ref{fig.6}(\textcolor[rgb]{0.98,0.00,0.00}{a})
and \ref{fig.6}(\textcolor[rgb]{0.98,0.00,0.00}{c}), respectively,
but with $\delta=0.1~\gamma_1$. One notices that, at normal
incidence and for $\delta=0.0~\gamma_1$ the transmission
probability shown in Figure
\ref{fig.6}(\textcolor[rgb]{0.98,0.00,0.00}{a}) is zero and there
are no resonances within a range of energy less than the height of
the barrier potential, i.e $E < V$. On the other hand resonances are present
at non-normal incidence as shown in Figure
\ref{fig.6}(\textcolor[rgb]{0.98,0.00,0.00}{c}). When the energy
is more than the height of the barrier potential the
transmission exhibits resonances. As observed in Figures
\ref{fig.6}(\textcolor[rgb]{0.98,0.00,0.00}{b}) and
\ref{fig.6}(\textcolor[rgb]{0.98,0.00,0.00}{d}) the transmission
probability is related to the transmission gap $\Delta
E=2~\delta$ and remains invariant for $E > V+\delta$. We also
observe that the number of resonances in the transmission as shown
in {Figures \ref{fig.6}(\textcolor[rgb]{0.98,0.00,0.00}{b})
and \ref{fig.6}(\textcolor[rgb]{0.98,0.00,0.00}{d})
decreases for $E < V-\delta$.}\\

\begin{figure}[H]
\centering
\includegraphics[width=5.5cm, height=4cm]{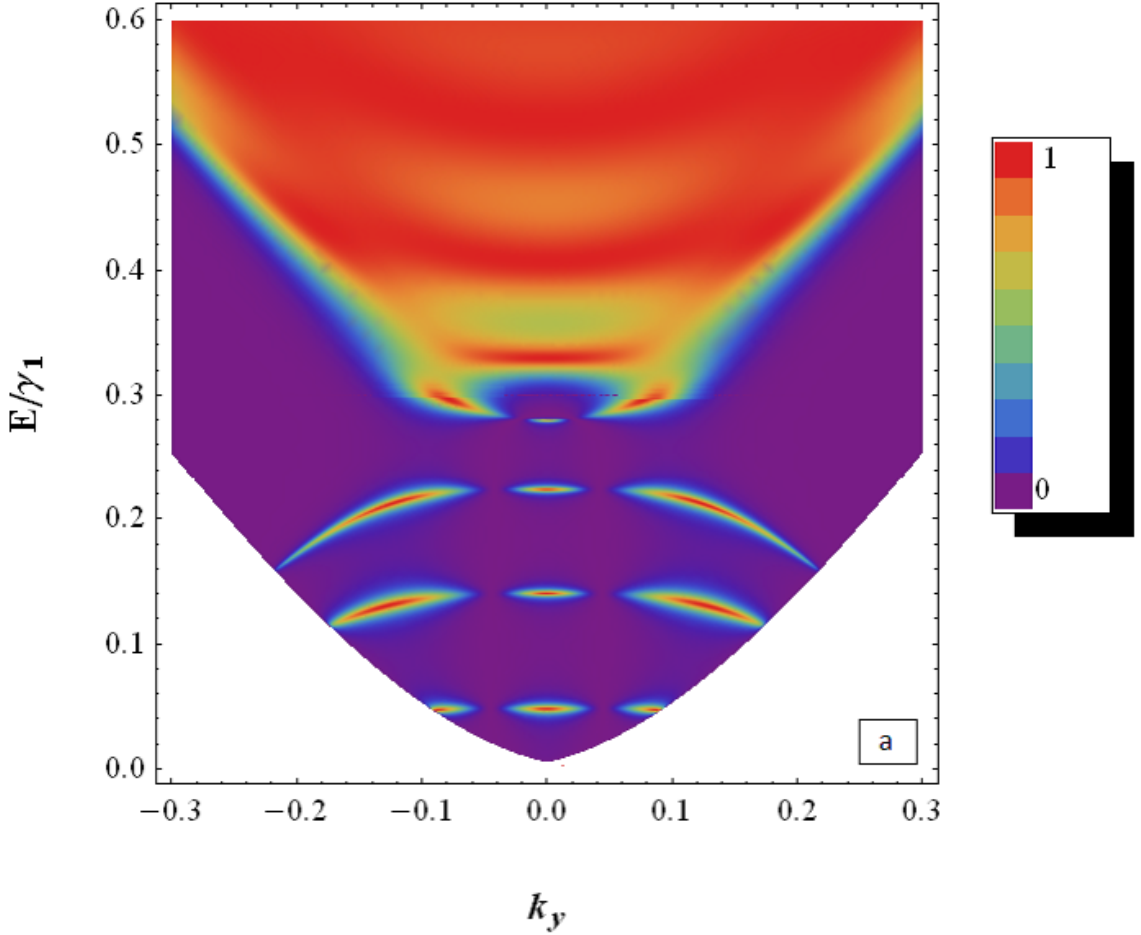}
\ \
\includegraphics[width=5.5cm, height=4cm]{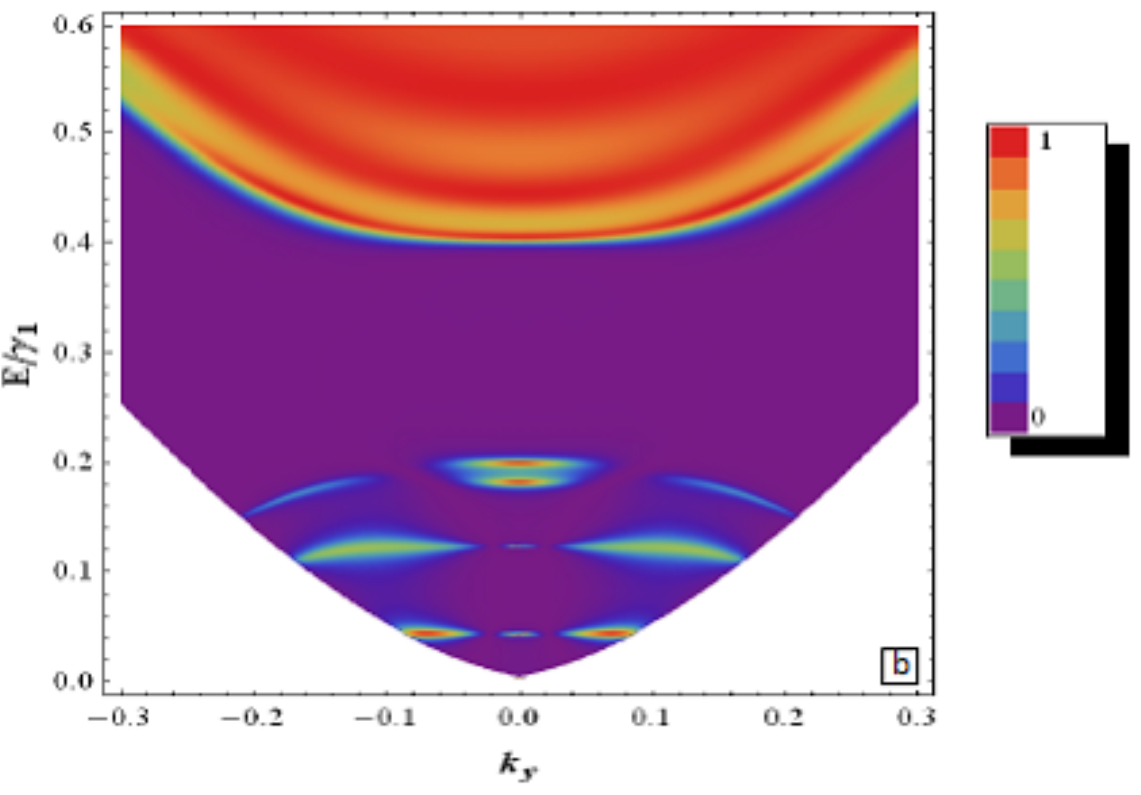}
\\
\includegraphics[width=5.5cm, height=4cm]{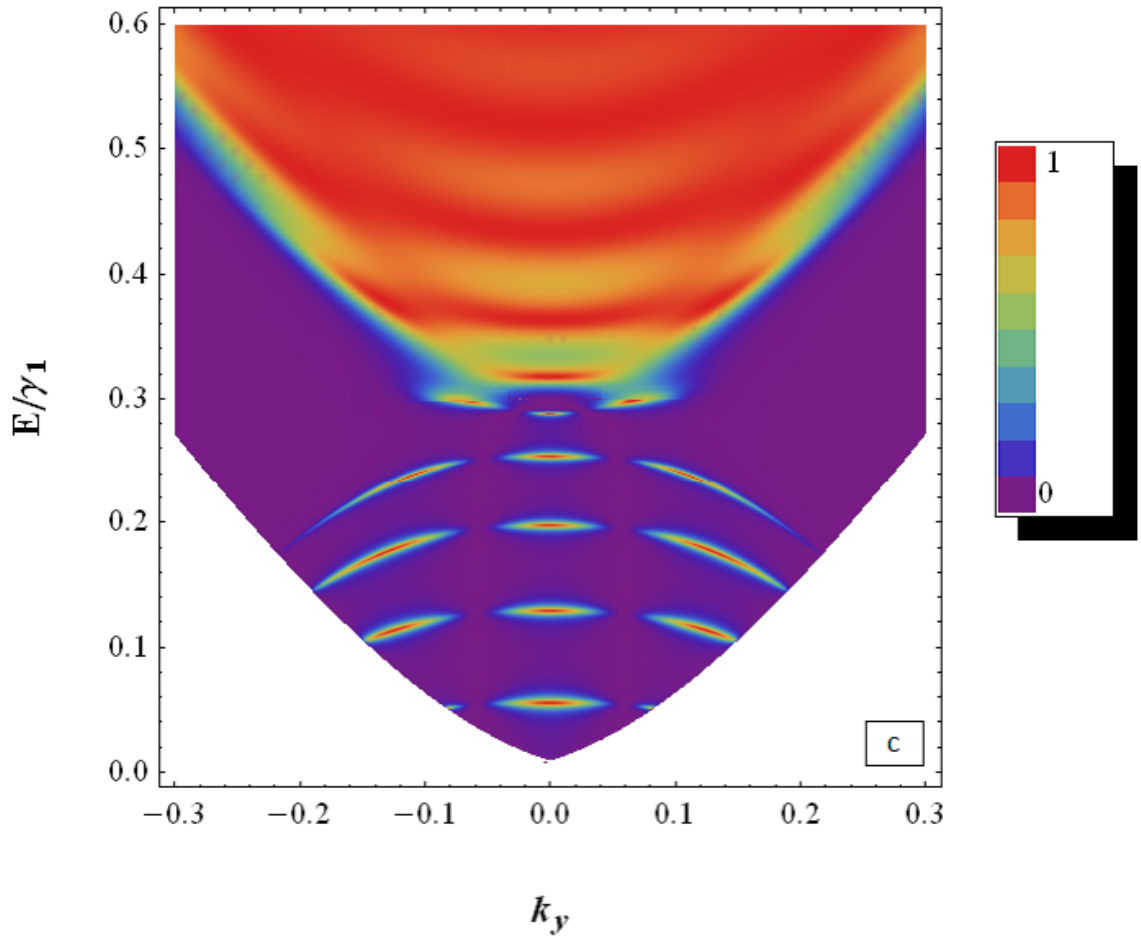}
\ \
\includegraphics[width=5.5cm, height=4cm]{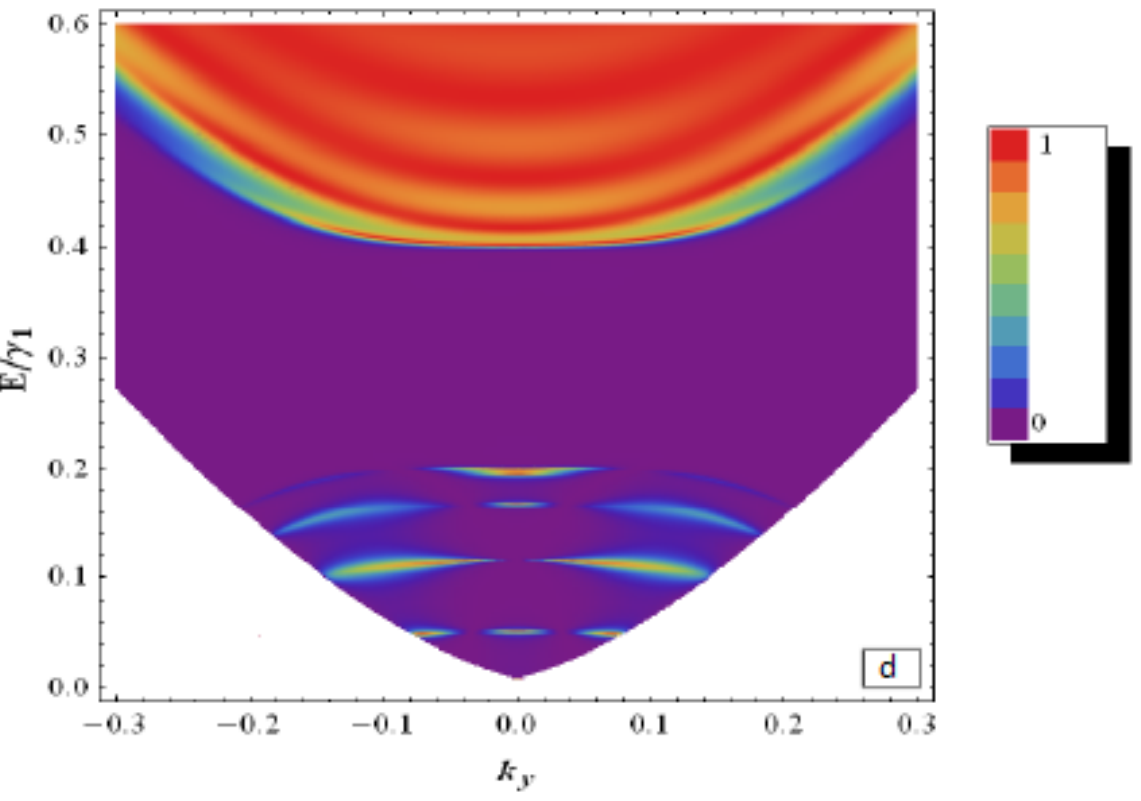}
\caption{Density plot of transmission probability as a function of
the transverse wave vector $k_y$ and energy $E$, for $V=0.3~
\gamma_1$ and $l_B=18.5~nm$.
(\textcolor[rgb]{0.98,0.00,0.00}{a})/(\textcolor[rgb]{0.98,0.00,0.00}{b})
for $\delta=0.0~\gamma_1$/$\delta=0.1~\gamma_1$, respectively, and
$d=30~nm$.
(\textcolor[rgb]{0.98,0.00,0.00}{c})/(\textcolor[rgb]{0.98,0.00,0.00}{d})
for $\delta=0.0~\gamma_1$/$\delta=0.1 ~\gamma_1$, respectively,
and $d=40~nm $.}\label{fig.7}
\end{figure}

In Figure \ref{fig.7}, we show the density plot of the transmission
probability as a function of the transverse wave vector $k_y$ and
energy $E$ for two values of the barrier width : $d=30~nm$
($d_2=-d_1=15~nm$) in Figures
\ref{fig.7}(\textcolor[rgb]{0.98,0.00,0.00}{a}) and
\ref{fig.7}(\textcolor[rgb]{0.98,0.00,0.00}{b}), and $d=40~nm$
($d_2=-d_1=20~nm$) in Figures
\ref{fig.7}(\textcolor[rgb]{0.98,0.00,0.00}{c}) and
\ref{fig.7}(\textcolor[rgb]{0.98,0.00,0.00}{d}). To see the effect
of the barrier width $d$ on the transmission probability  at
non-normal incidence we show in Figure
\ref{fig.7}(\textcolor[rgb]{0.98,0.00,0.00}{a}) that when we increase $d$ a new peak of
resonance appear within the range of energy less than the height of the barrier potential, i.e $E < V$.
The number of these resonance peaks depends on the width of the well
between the barriers. At nearly normal incidence $\mid k_y \mid
\approx -\mid \frac{d_1}{l_{B}^2} \mid\approx 0.04~nm^{-1}$ in
Figure \ref{fig.7}(\textcolor[rgb]{0.98,0.00,0.00}{a}), and $\mid
k_y \mid \approx -\mid \frac{d_1}{l_{B}^2} \mid \approx 0.06
~nm^{-1}$ in Figure
\ref{fig.7}(\textcolor[rgb]{0.98,0.00,0.00}{c}) we have zero
transmission when the energy is less than the height of the
barrier potential. On the other hand, for energy more than the
height of the barrier the Dirac fermions exhibit transmission
resonances as seen in Figures
\ref{fig.7}(\textcolor[rgb]{0.98,0.00,0.00}{a}) and
\ref{fig.7}(\textcolor[rgb]{0.98,0.00,0.00}{b}), the
number of transmission resonances increase when we increase the
barrier width $d$ as shown in Figures
\ref{fig.7}(\textcolor[rgb]{0.98,0.00,0.00}{c}) and
\ref{fig.7}(\textcolor[rgb]{0.98,0.00,0.00}{d}), respectively. We
remark from Figures
\ref{fig.7}(\textcolor[rgb]{0.98,0.00,0.00}{b}) and
\ref{fig.7}(\textcolor[rgb]{0.98,0.00,0.00}{d}) that the transmission
probability is correlated to the transmission gap $\Delta E = 2~\delta$.

\begin{figure}[H]
\centering
\includegraphics[width=6.5cm, height=4.5cm]{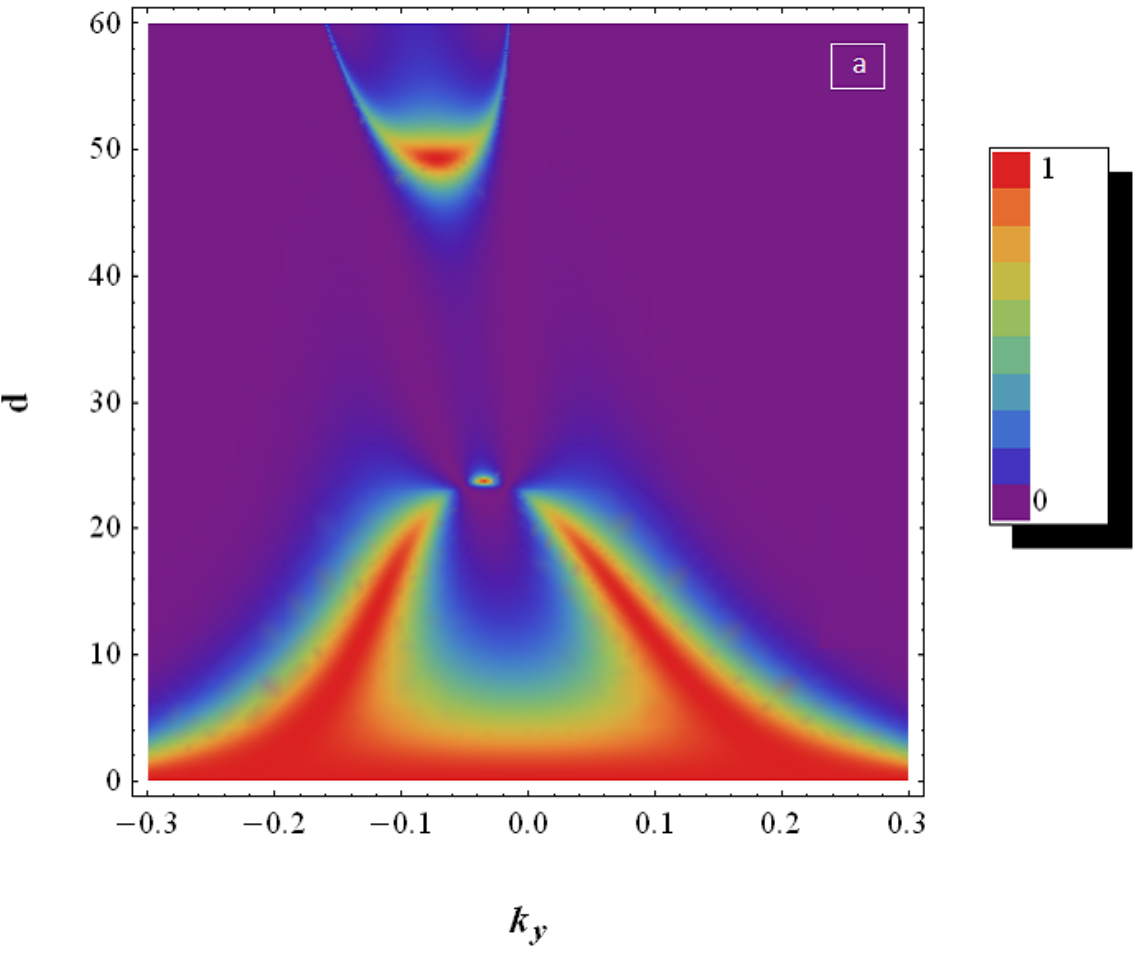}
\ \
\includegraphics[width=6.5cm, height=4.5cm]{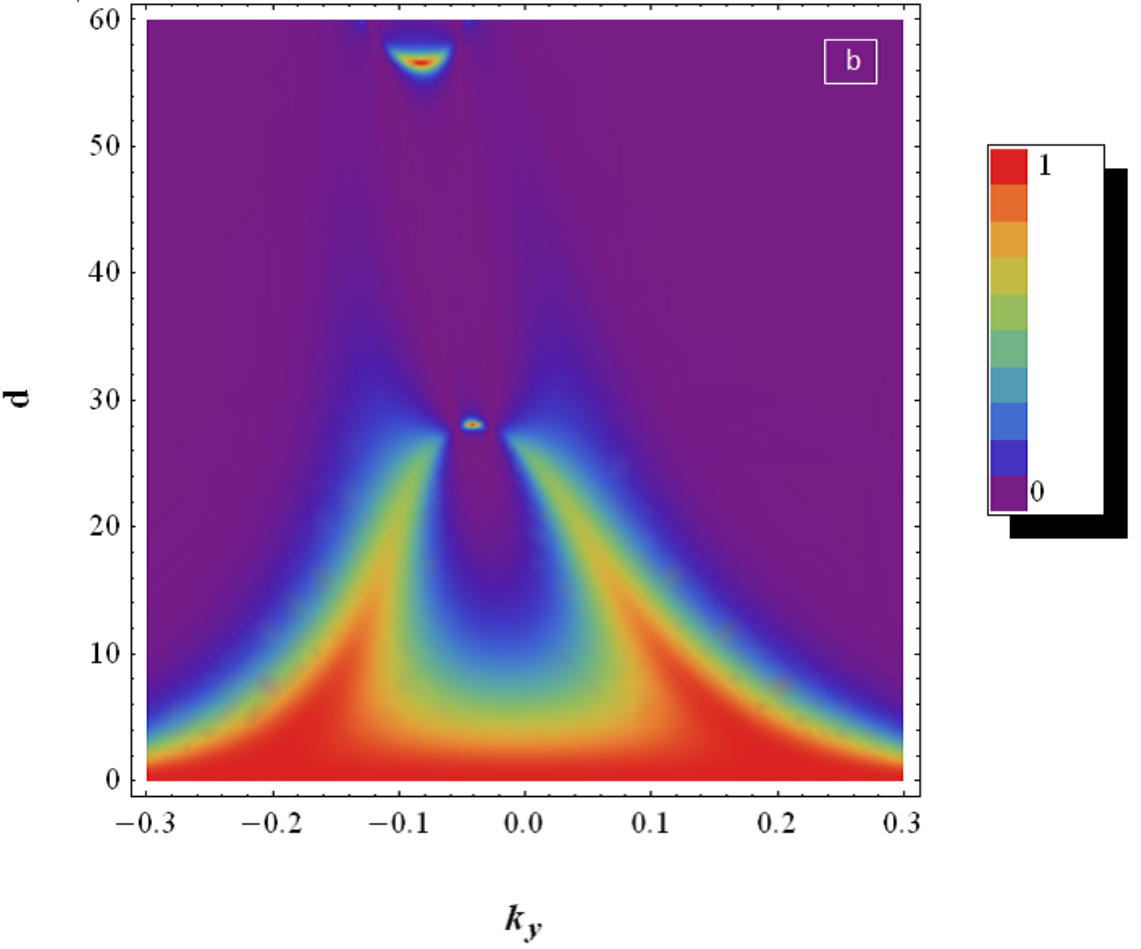}
\\
\includegraphics[width=6.5cm, height=4.5cm]{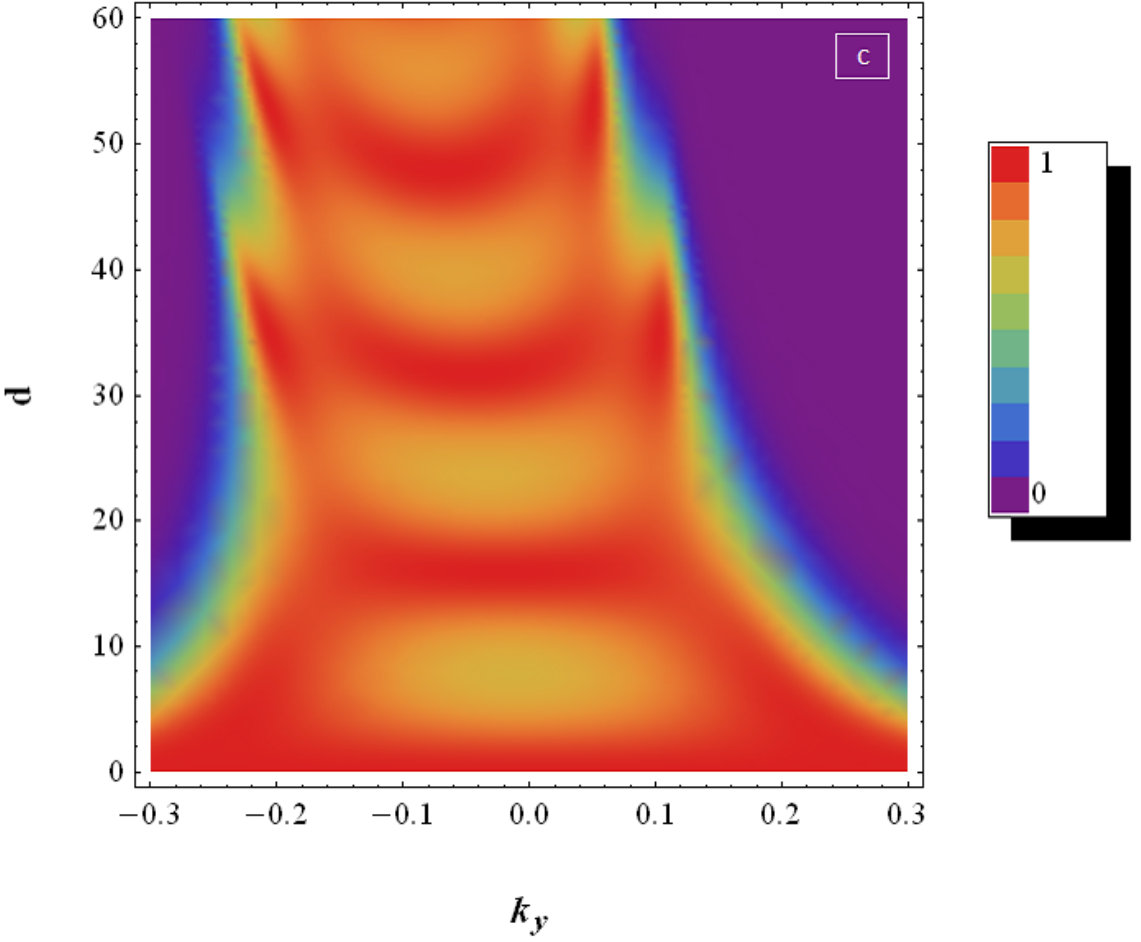}
\ \
\includegraphics[width=6.5cm, height=4.5cm]{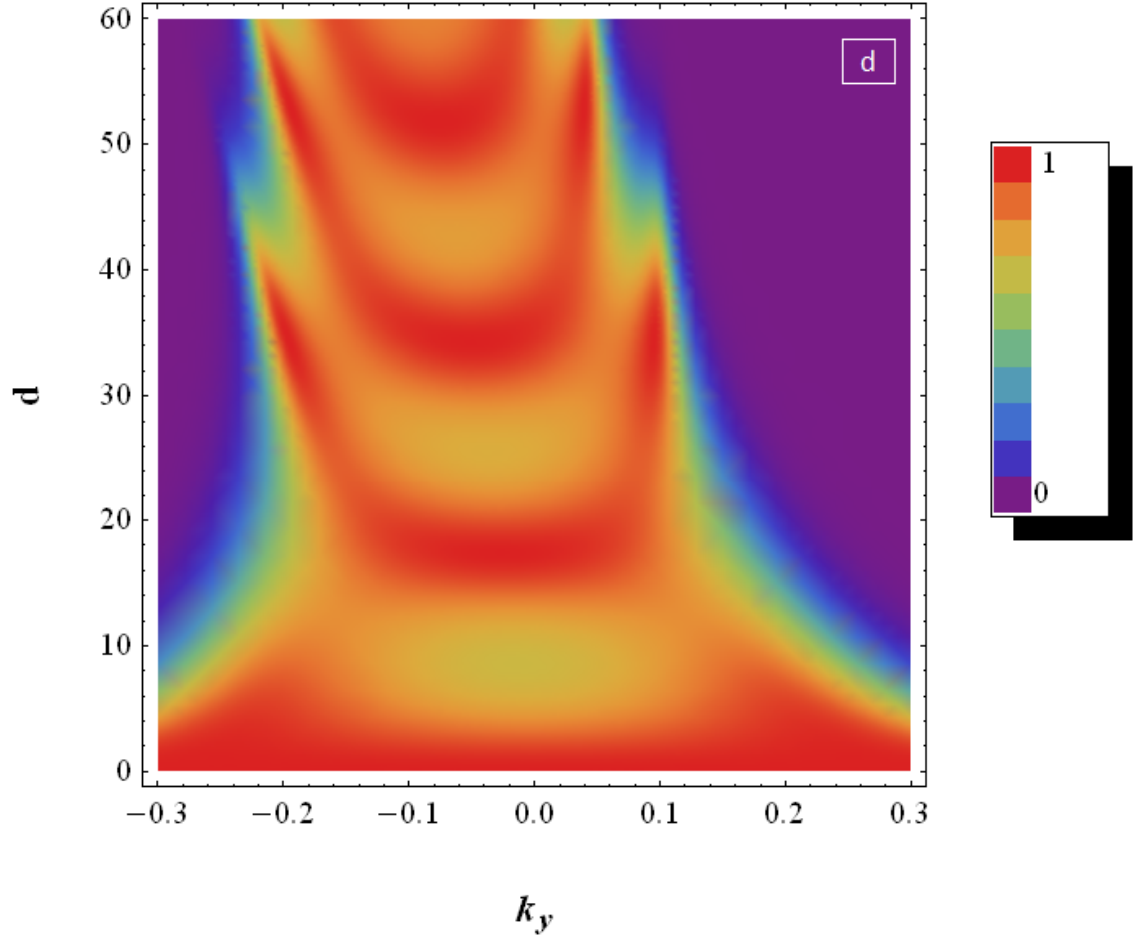}
\caption{Density plot of transmission probability as a function of
the transverse wave vector $k_y$ and the barrier width $d$,
for $V=0.3~\gamma_1$ and $l_B=18.5~nm$.
(\textcolor[rgb]{0.98,0.00,0.00}{a})/(\textcolor[rgb]{0.98,0.00,0.00}{b})
for $\delta=0.0~\gamma_1$/$\delta=0.02~\gamma_1$ and
$E=\frac{9}{10}~V$.
(\textcolor[rgb]{0.98,0.00,0.00}{c})/(\textcolor[rgb]{0.98,0.00,0.00}{d})
for $\delta=0.0~\gamma_1$/$\delta=0.05~\gamma_1$ and
$E=\frac{13}{10}~V$.}\label{fig.8}
\end{figure}

In Figure \ref{fig.8} we show the density plot of transmission
probability as function of the transfer wave vector $k_y$ and the
barrier width $d$, for $V=0.3~\gamma_1$ and $l_B=18.5~nm$. In
Figures \ref{fig.8}(\textcolor[rgb]{0.98,0.00,0.00}{a}) and
\ref{fig.8}(\textcolor[rgb]{0.98,0.00,0.00}{b}) we fix the energy
at $E=\frac{9}{10}~V$, for two different values of the interlayer
potential $\delta=0.0~\gamma_1$ and $\delta=0.02~\gamma_1$. In
Figures \ref{fig.8}(\textcolor[rgb]{0.98,0.00,0.00}{c}) and
\ref{fig.8}(\textcolor[rgb]{0.98,0.00,0.00}{d}) we fix the energy
at $E=\frac{13}{10}~V$ again for two different values of the
interlayer electrostatic potential $\delta=0.0~\gamma_1$ and
$\delta=0.05~\gamma_1$. For $\delta=0.0~\gamma_1$ and for energy
less than the height of the potential barrier, $E < V$, we have
full transmission for a wide range of $k_y$ values. {By increasing} the
width $d$, we create one resonance peak as depicted in Figure
\ref{fig.8}(\textcolor[rgb]{0.98,0.00,0.00}{a}). However, the
total transmission probability decreases for
$\delta=0.02~\gamma_1$ as shown in Figure
\ref{fig.8}(\textcolor[rgb]{0.98,0.00,0.00}{b}). In Figure
\ref{fig.8}(\textcolor[rgb]{0.98,0.00,0.00}{c}) most of the
resonances disappear while oscillations take over in the
transmission.  {The number} of oscillations decrease in presence of
the interlayer electrostatic potential as reflected in Figure
\ref{fig.8}(\textcolor[rgb]{0.98,0.00,0.00}{d}).

\section{Four band tunneling}

Once we allow for higher energies, $E > \gamma_1$, we will have
four transmission and four reflection channels resulting in what we call the
four band tunneling.\\

\begin{figure}[h!]
\centering
\includegraphics[width=5.5cm, height=4cm]{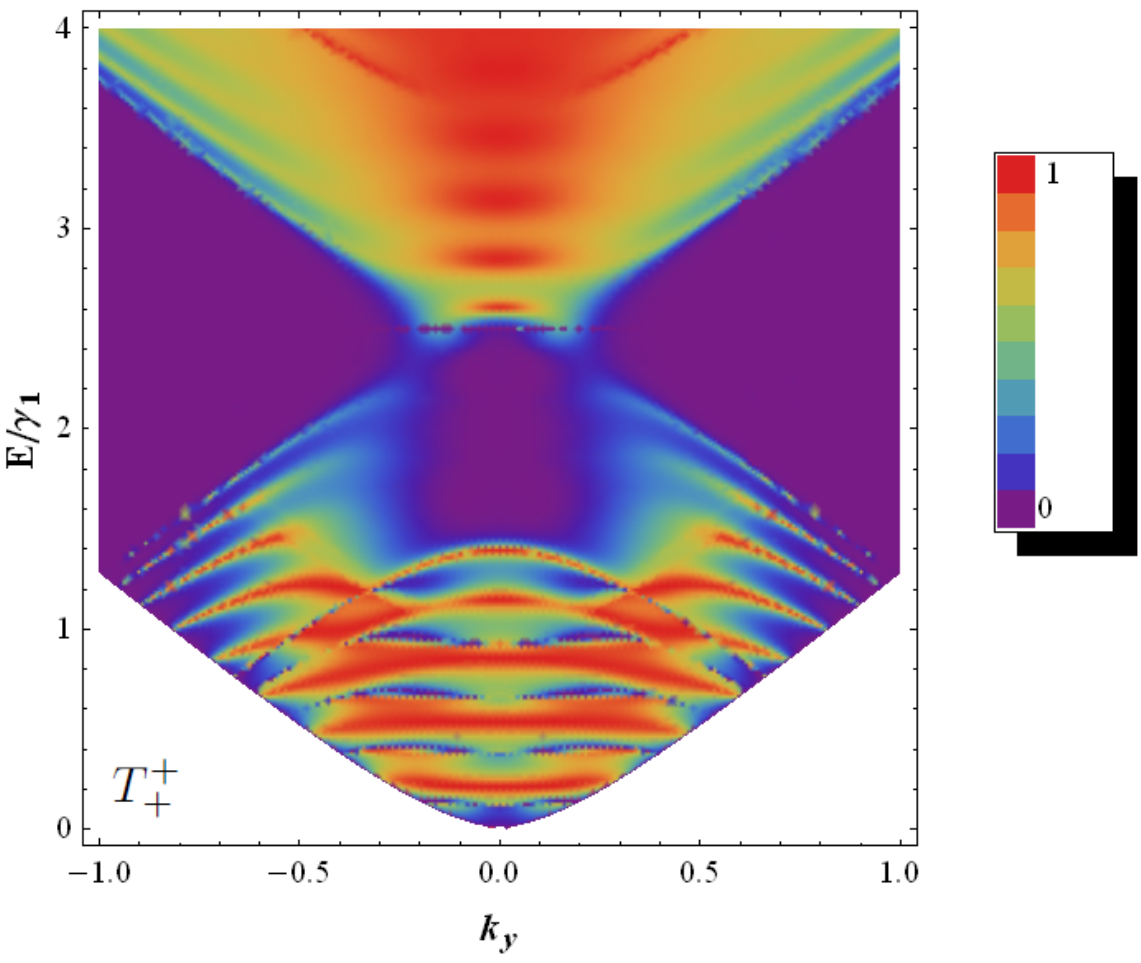}
\
\includegraphics[width=5.5cm, height=4cm]{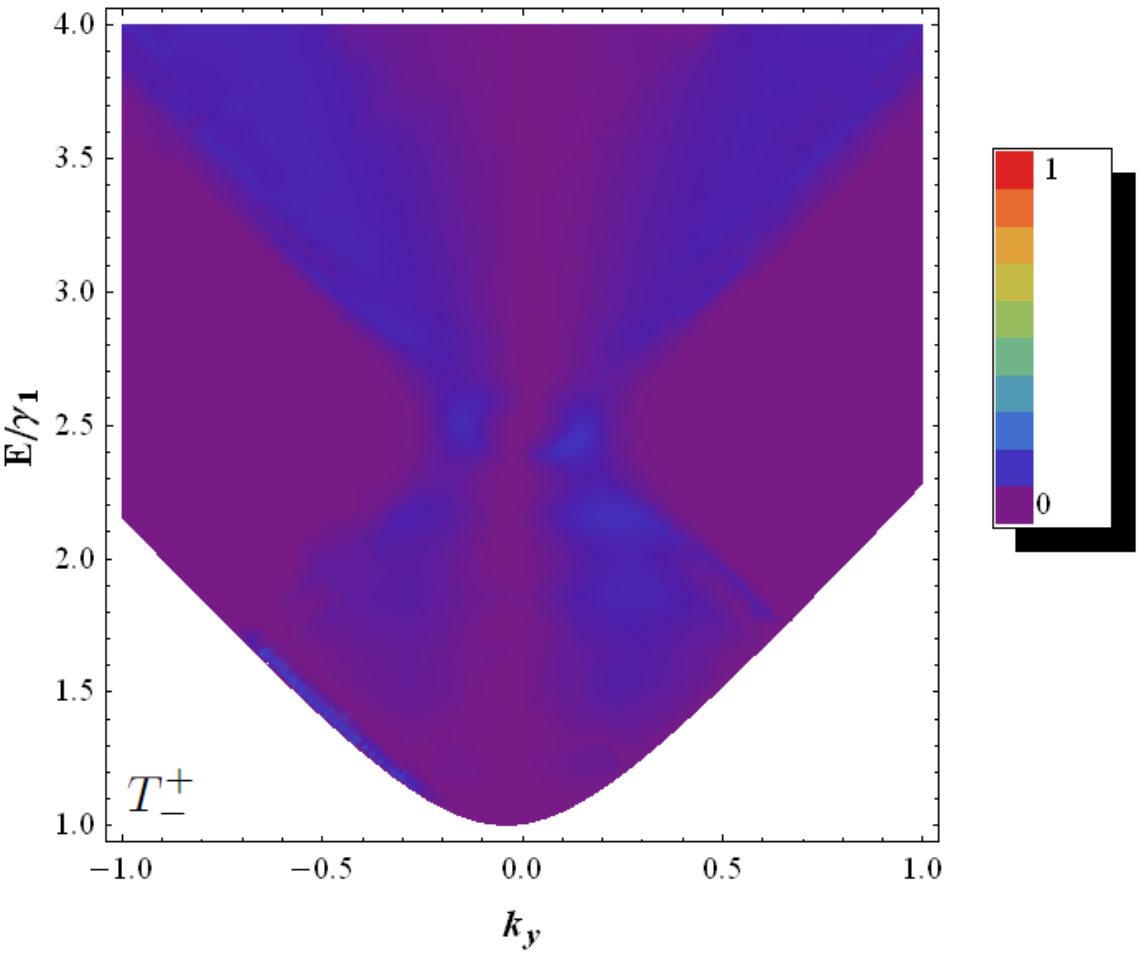}
\
\includegraphics[width=5.5cm, height=4cm]{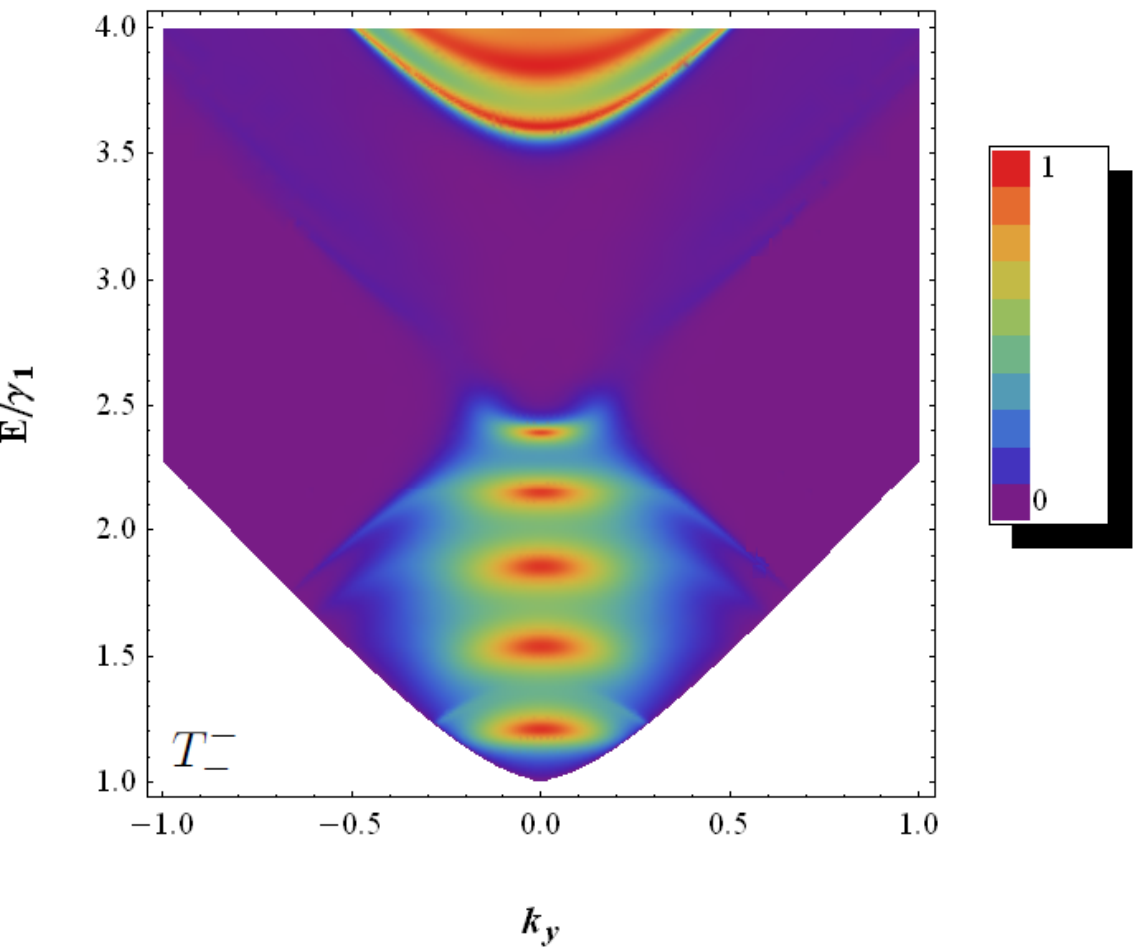}
\\
\includegraphics[width=5.5cm, height=4cm]{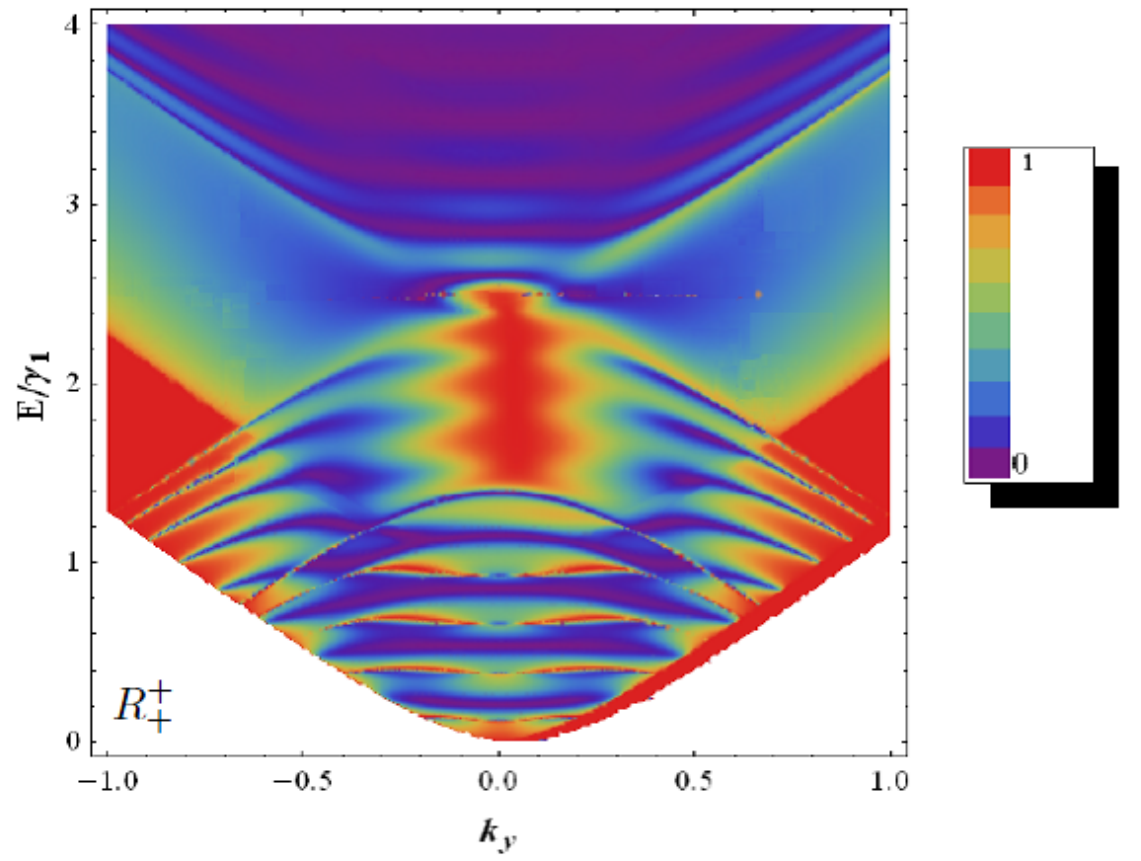}
\
\includegraphics[width=5.5cm, height=4cm]{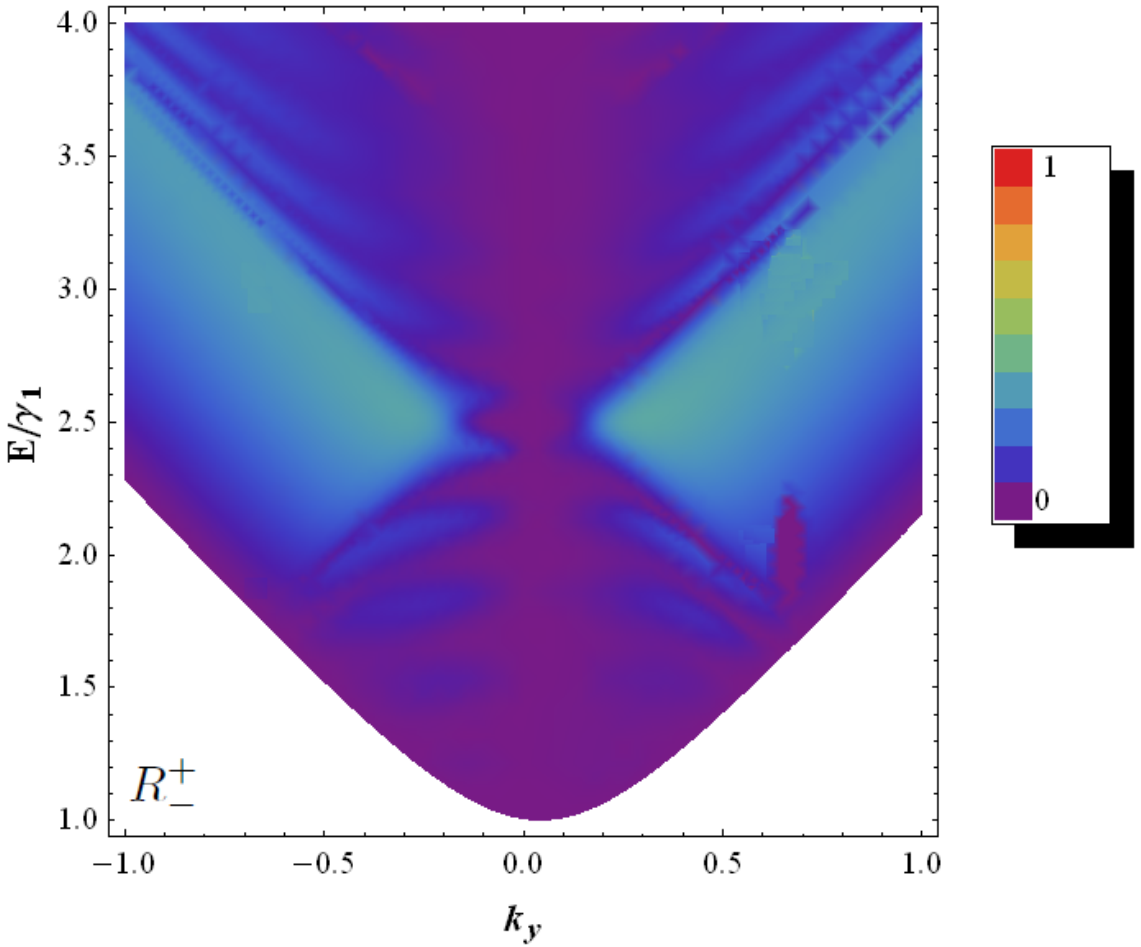}
\
\includegraphics[width=5.5cm, height=4cm]{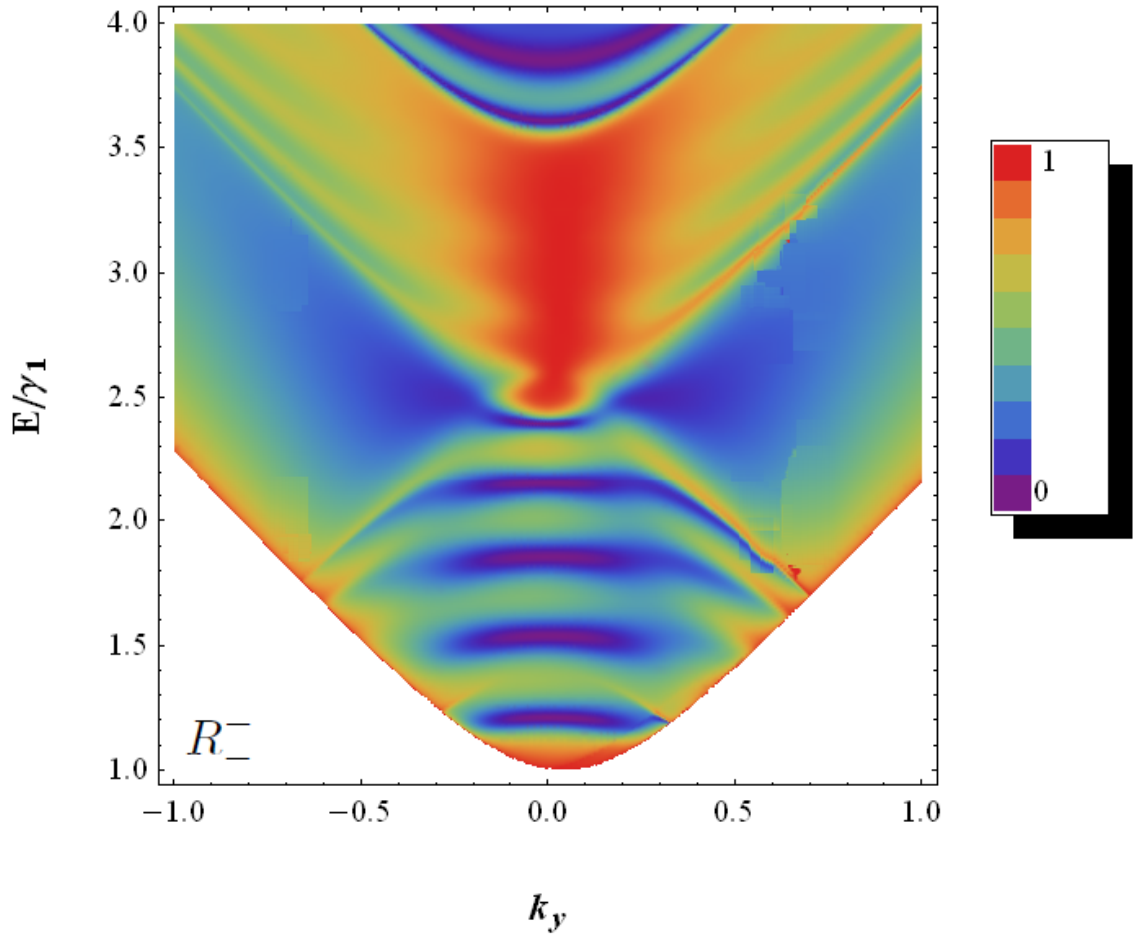}
\\
\includegraphics[width=5.5cm, height=4cm]{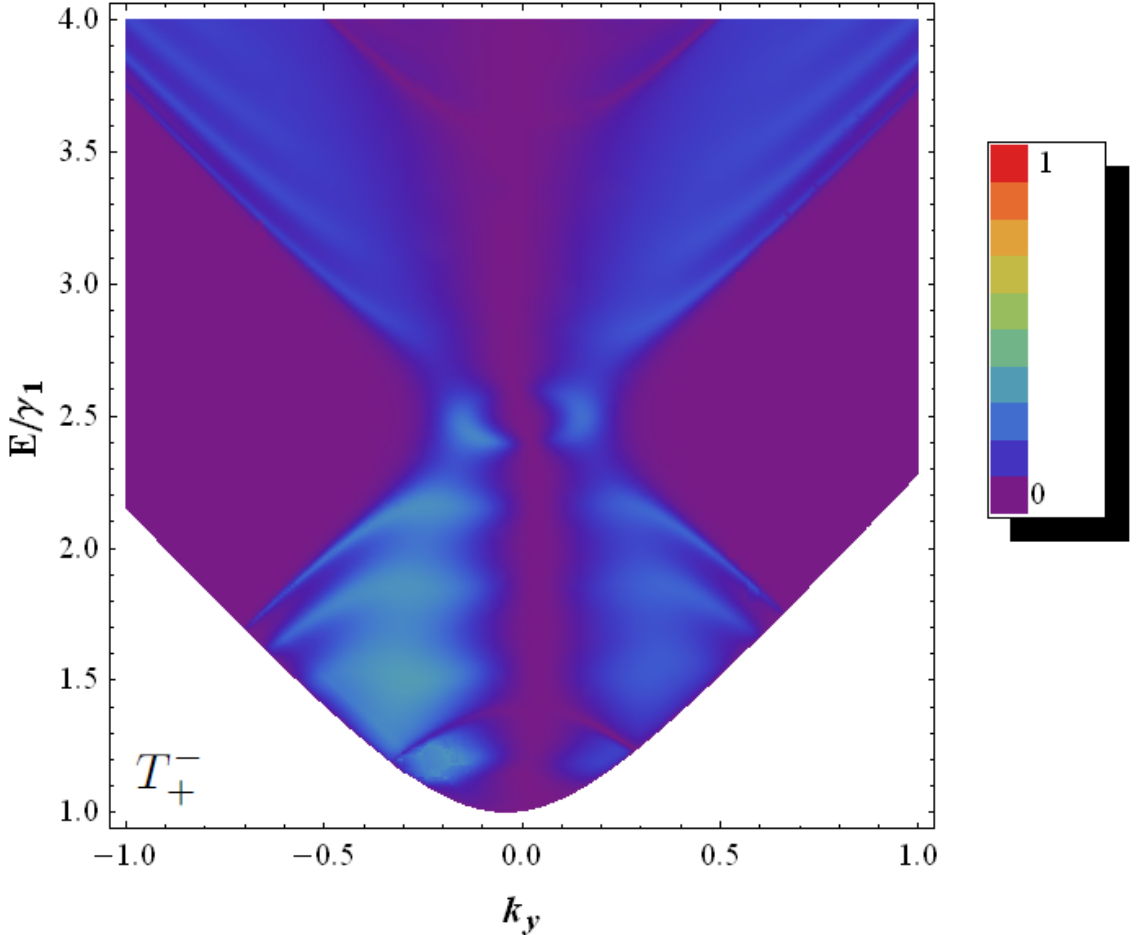}
\
\includegraphics[width=5.5cm, height=4cm]{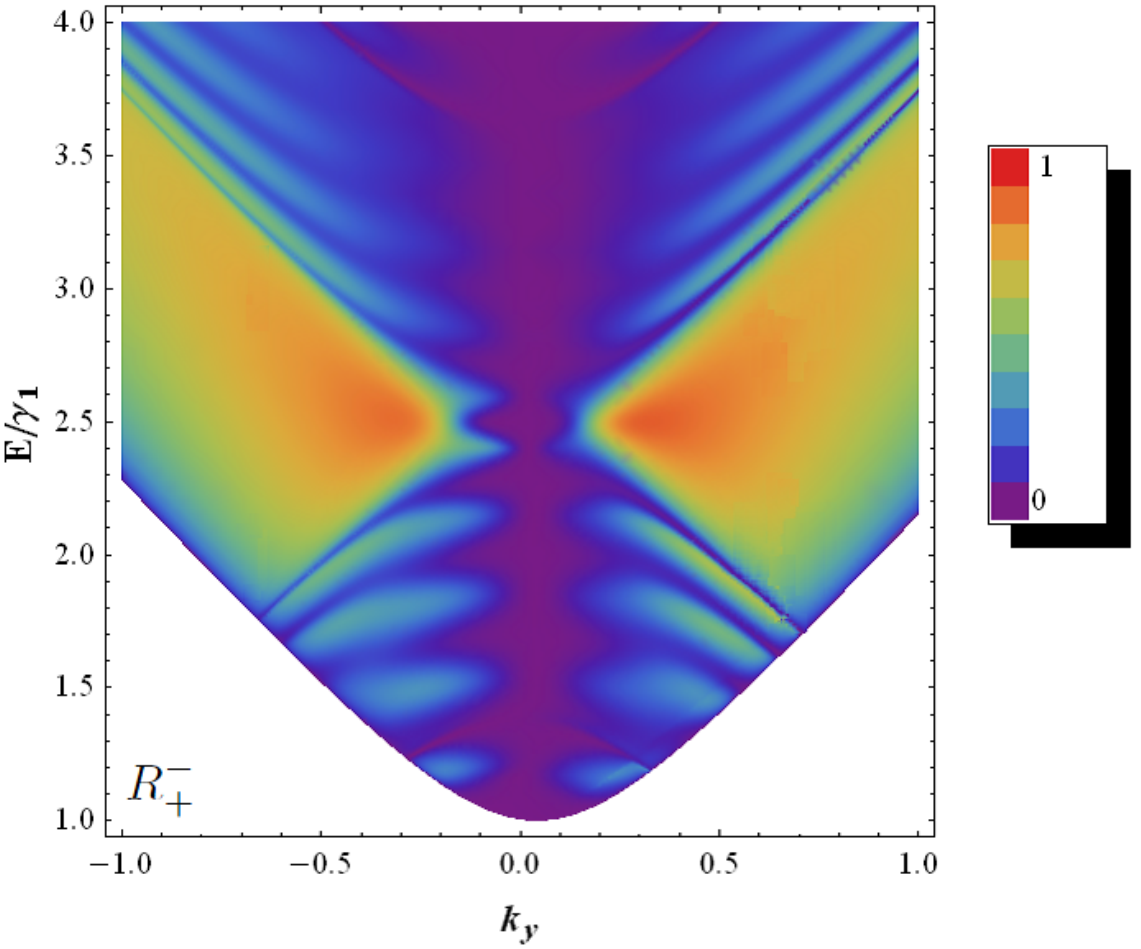}
\caption{Density plot of transmission and reflection coefficients
as a function of the transverse wave vector $k_y$ and energy
$E$ with $V=2.5~\gamma_1$, $\delta=0.0~\gamma_1$, $l_B=13.5~nm$ ,
and $d_2=-d_1=7.5~nm$ .}\label{fig.9}
\end{figure}

In Figure \ref{fig.9} we show the
transmission and reflection {probabilities} associated with different channels,
as a function of the transverse wave vector $k_y$ and the incident energy $E$,
we used $V=2.5~\gamma_1$, $\delta=0.0~\gamma_1$, $l_B=13.5~nm$, and $d_2=-d_1=7.5~nm$.
For energies less than $V-\gamma_1$ the Dirac fermions exhibit
transmission resonances in $T_{+}^{+}$ in which the electrons
propagate via $\alpha^+$ mode inside the barriers. For $V-\gamma_1
< E < V$, there are no available $\alpha^+$ states and the
transmission is suppressed in this region. For nearly normal
incidence, $k_y \approx -\frac{d_2}{l_{B}^2} \approx
-0.04~nm^{-1}$ for $T_{-}^{+}$ and $k_y \approx
-\frac{d_1}{l_{B}^2} \approx 0.04~nm^{-1}$ for $T_{+}^{-}$, the
cloak effect \cite{Gu} occurs in the energy region $V-\gamma_1 < E
< V$, where the two modes $\alpha^+$ outside and inside barrier
regions are decoupled and therefore no scattering occur between
them \cite{Duppen} in the $T_{-}^{+}$ and $T_{+}^{-}$ channels.
While for non-normal incidence the two modes $\alpha^+$ outside
and inside barrier region are coupled, so that the transmission
$T_{-}^{+}$ and $T_{+}^{-}$ channels in the same energy region are
non-zero. The transmission probabilities $T_{-}^{+}$ and
$T_{+}^{-}$ are different ( $T_{-}^{+} \neq T_{+}^{-}$), which
introduces an asymmetry for a single barrier due to the presence of the
magnetic field. In addition, the reflection coefficients $R_{-}^{+}$ and $R_{+}^{-}$
are different ($R_{-}^{+} \neq R_{+}^{-}$) and do not have the same number of
resonances and anti-resonance, these observations were absent in
the case of single and double barrier in the absence of magnetic
field \cite{Duppen,Hassan}. For $T_{-}^{-}$ and $R_{-}^{-}$ the electrons propagate via
$\alpha^-$ mode for $E < V$ and $ E > V+\gamma_1 $, which is
blocked inside the barrier for $V < E < V+\gamma_1$ so that the
transmission is suppressed in this region and this is equivalent
to the cloak effect \cite{Duppen,Hassan}.

\begin{figure}[H]
\centering
\includegraphics[width=5.5cm, height=4cm]{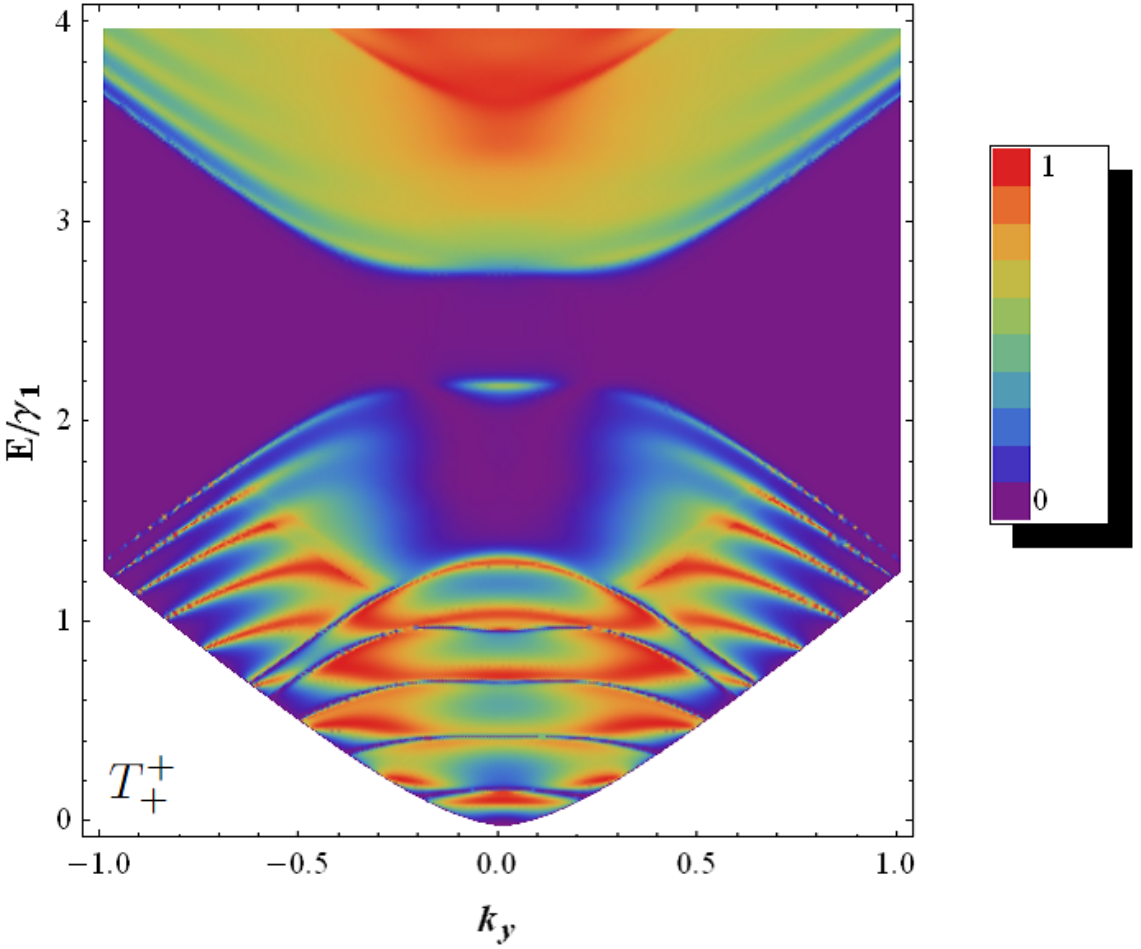}
\
\includegraphics[width=5.5cm, height=4cm]{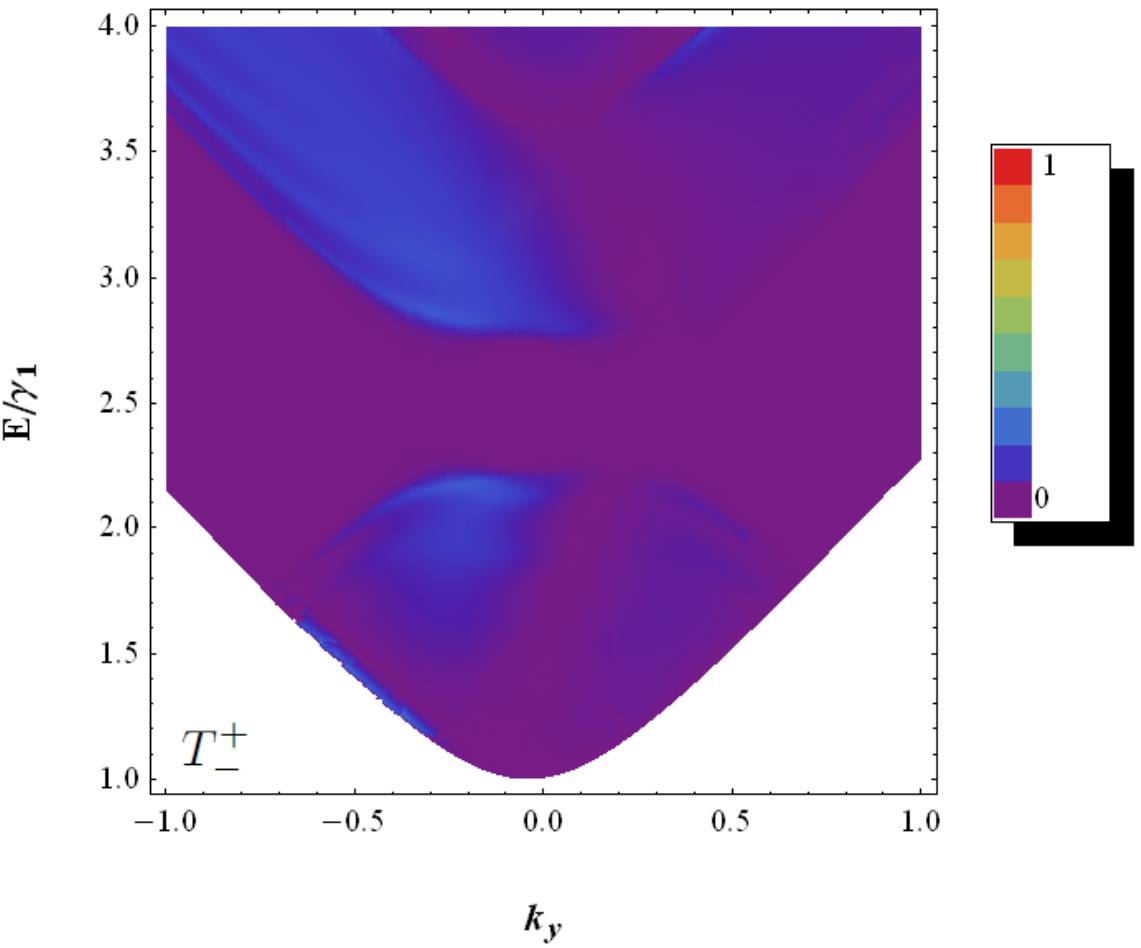}
\
\includegraphics[width=5.5cm, height=4cm]{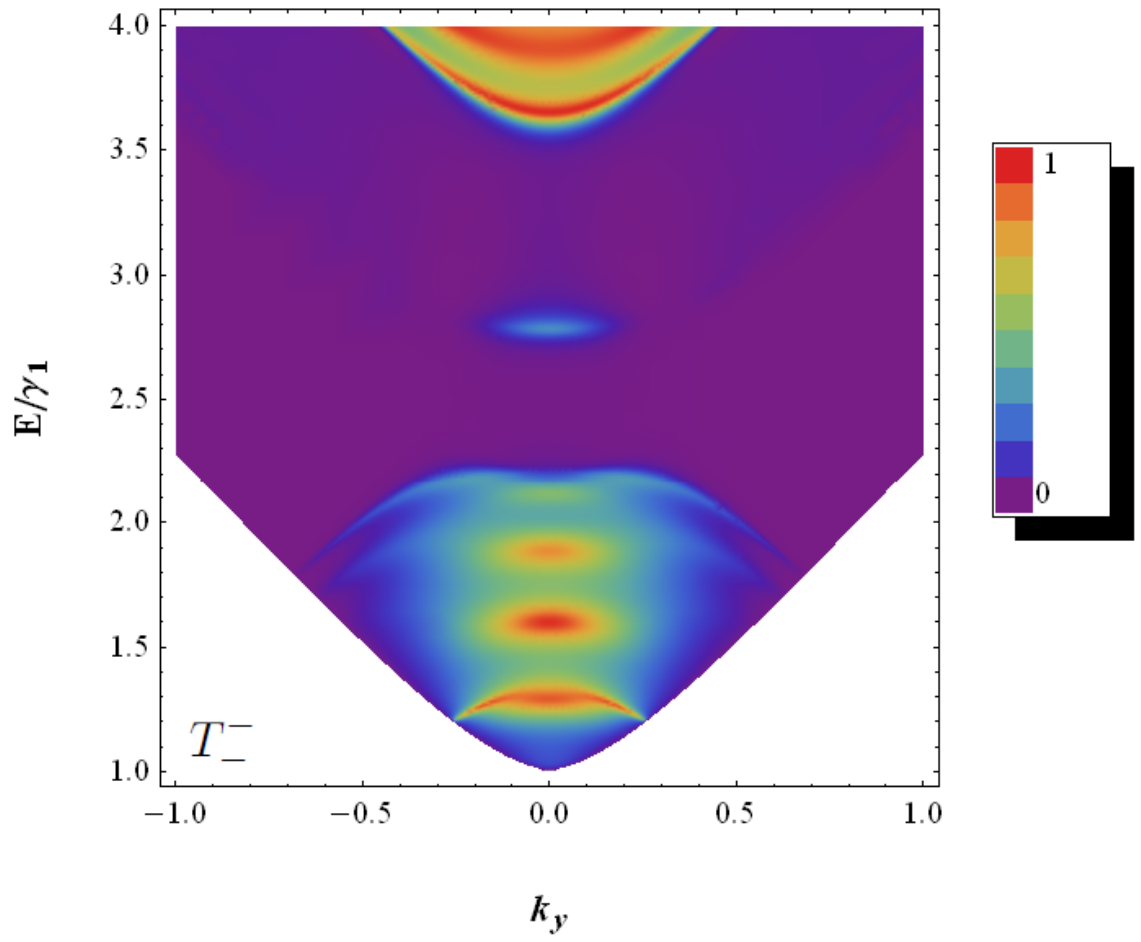}
\\
\includegraphics[width=5.5cm, height=4cm]{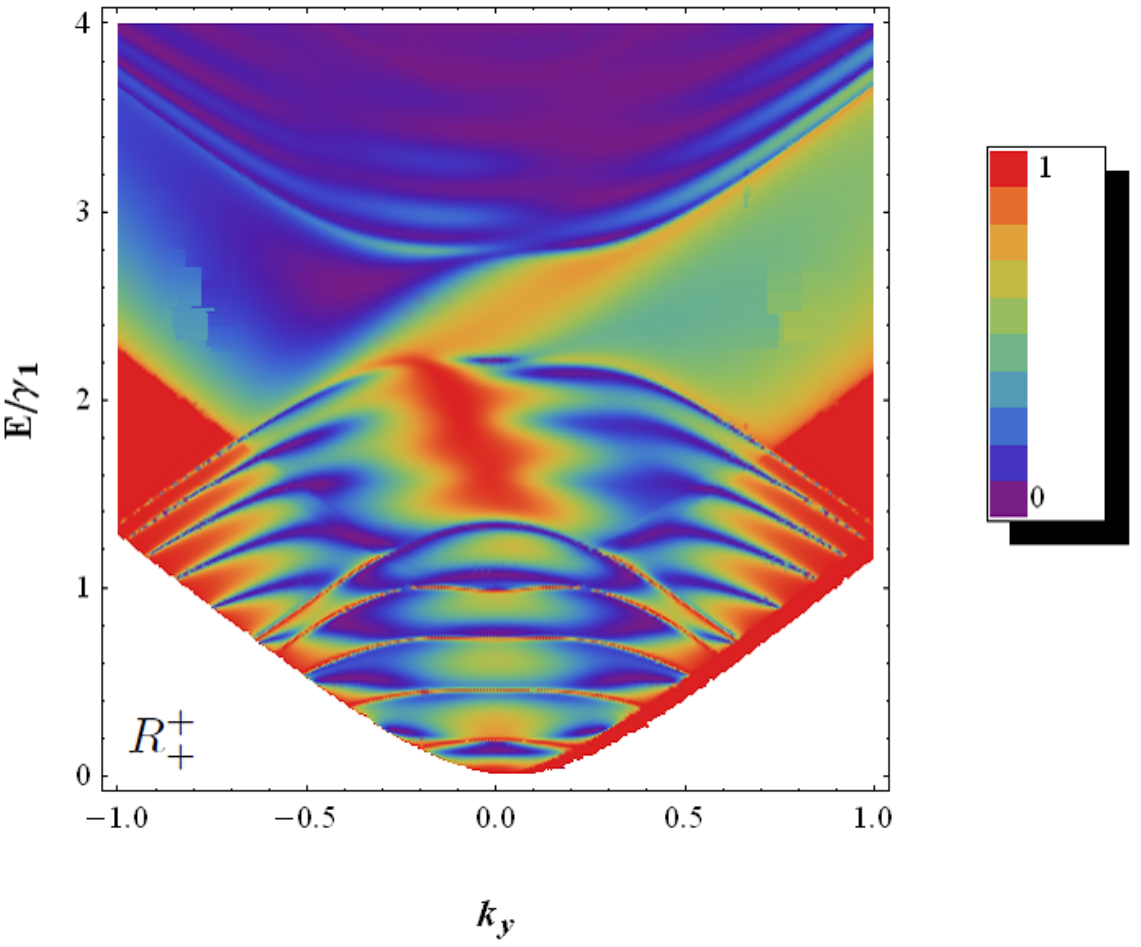}
\
\includegraphics[width=5.5cm, height=4cm]{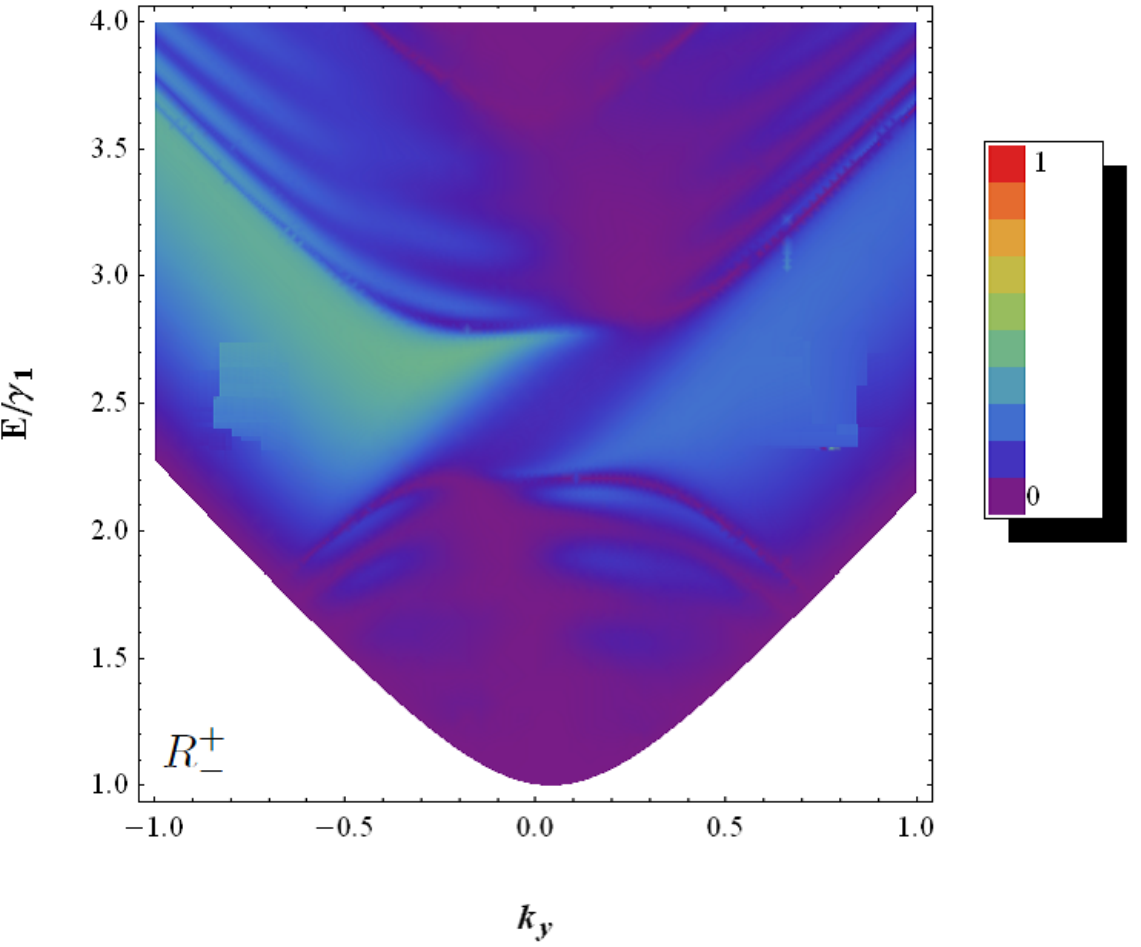}
\
\includegraphics[width=5.5cm, height=4cm]{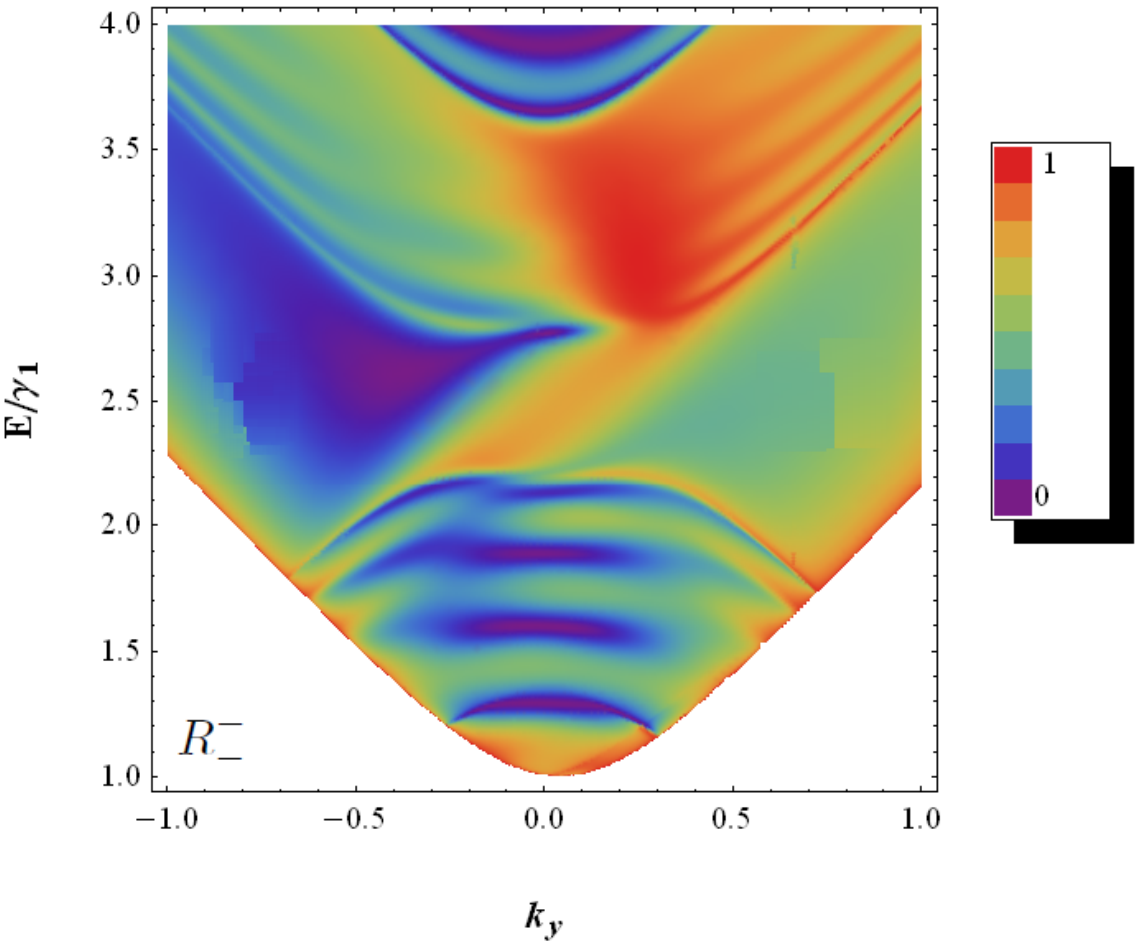}
\\
\includegraphics[width=5.5cm, height=4cm]{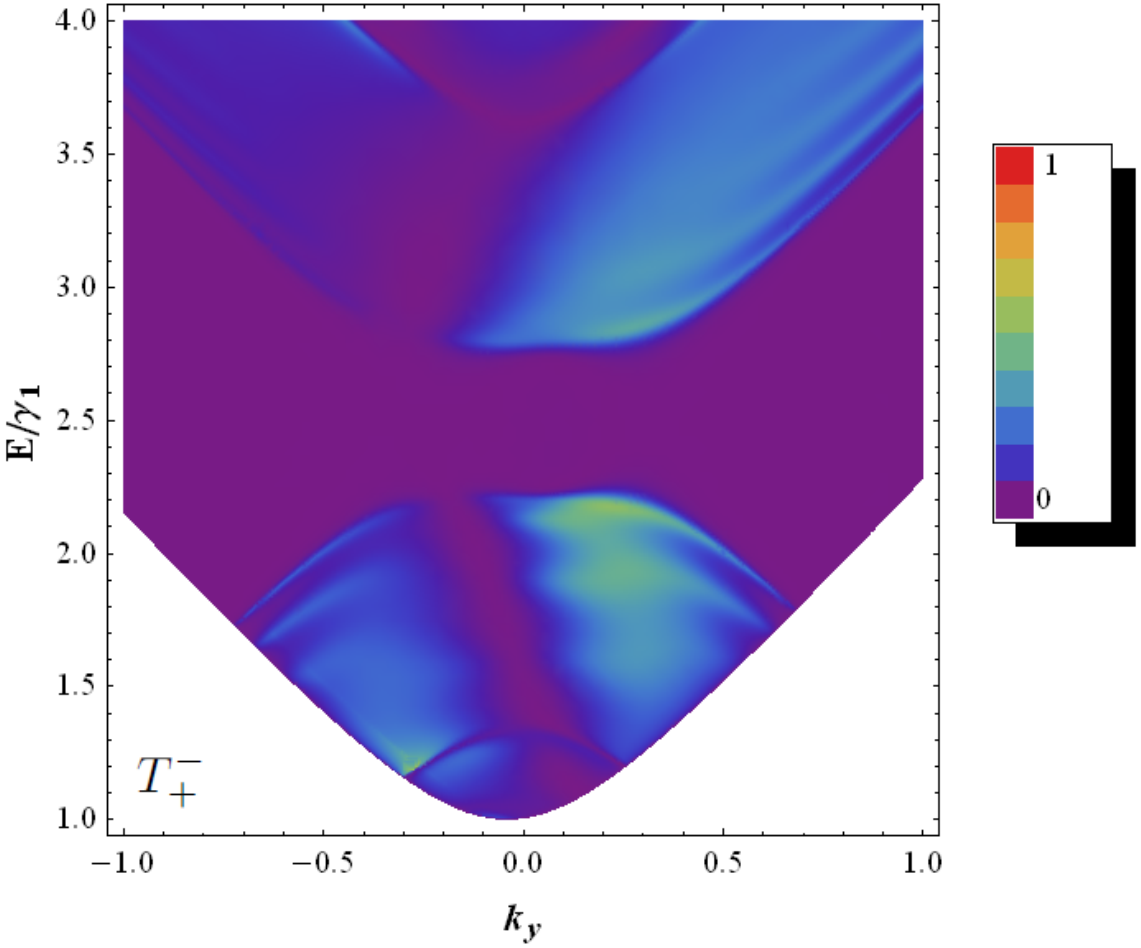}
\includegraphics[width=5.5cm, height=4cm]{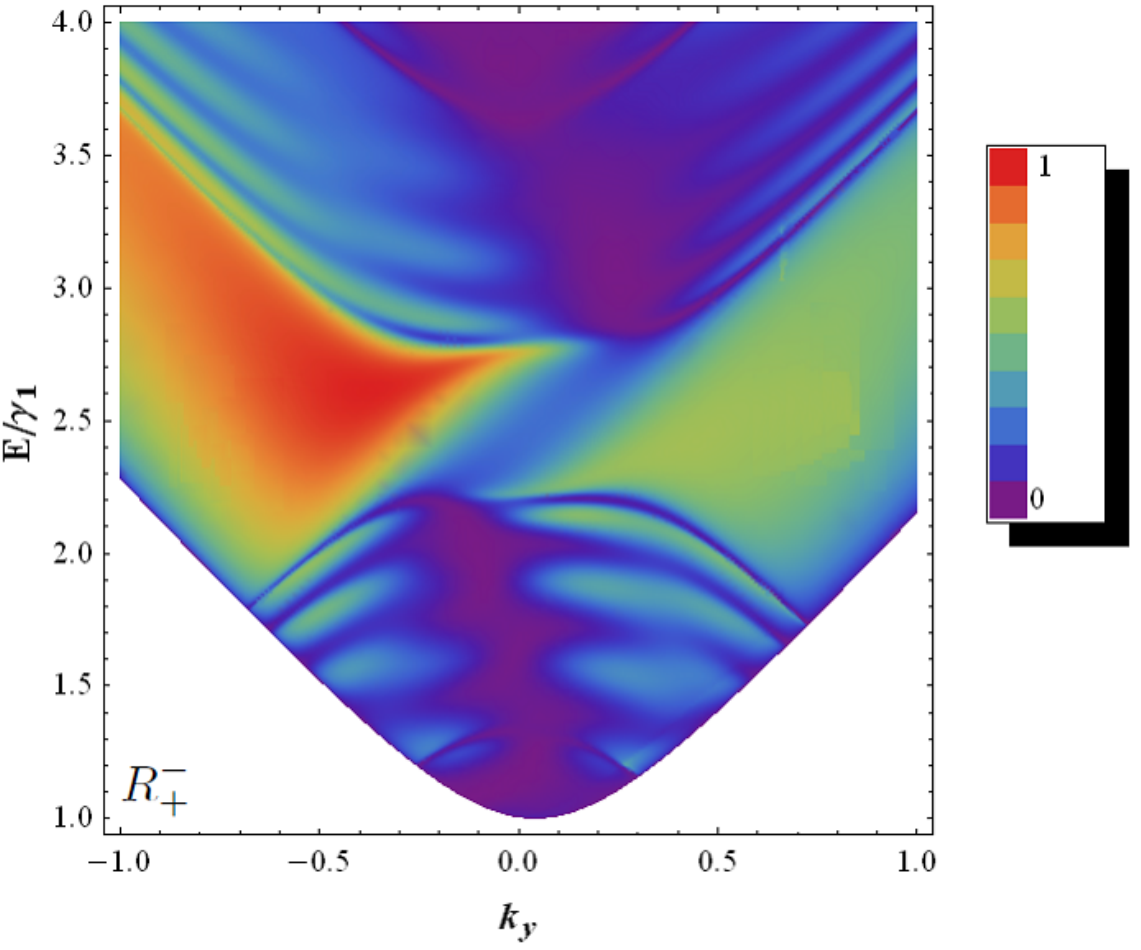}
\caption{Density plot of transmission and reflection coefficients
as a function of the transverse wave vector $k_y$ and energy
$E$ for $V=2.5~\gamma_1$, $\delta=0.3~\gamma_1$, $l_B=13.5~nm$,
and $d_2=-d_1=7.5~nm$.}\label{fig.10}
\end{figure}
To probe the effect of the interlayer electrostatic potential $\delta$, we
investigate the density plot of the transmission probability as
function of the transfer wave vector $k_y$ and energy $E$, using the
same parameters as in Figure \ref{fig.8} but for
$\delta=0.3~\gamma_1$ in Figure \ref{fig.10}. We note that the transmission probability
in the energy region $V-\delta < E < V+\delta$ is correlated to the
transmission gap and shows a suppression due to cloak effect, as it
was the case for the single barrier \cite{Duppen}.

\section{Conductance}

\begin{figure}[H]
\centering
\includegraphics[width=5.5cm, height=4cm]{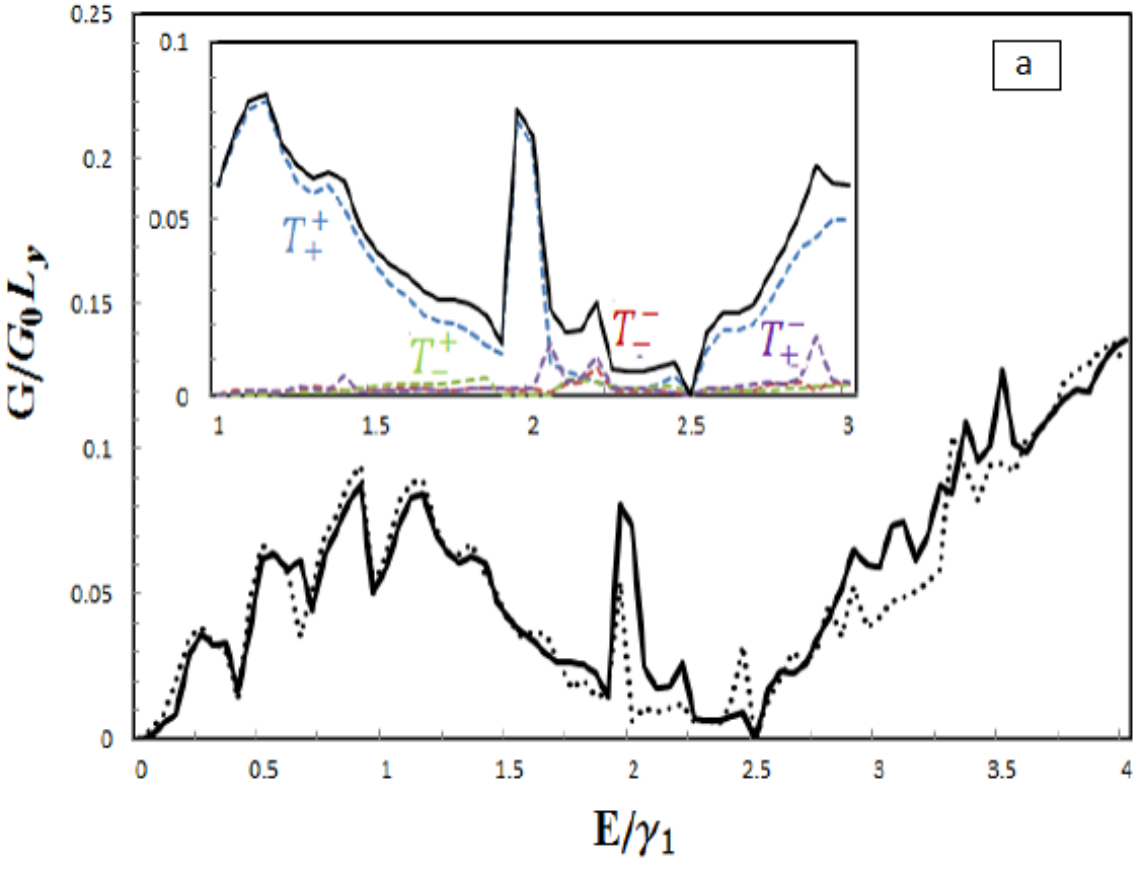}
\ \ \ \ \ \ \ \ \ \
\includegraphics[width=5.5cm, height=4cm]{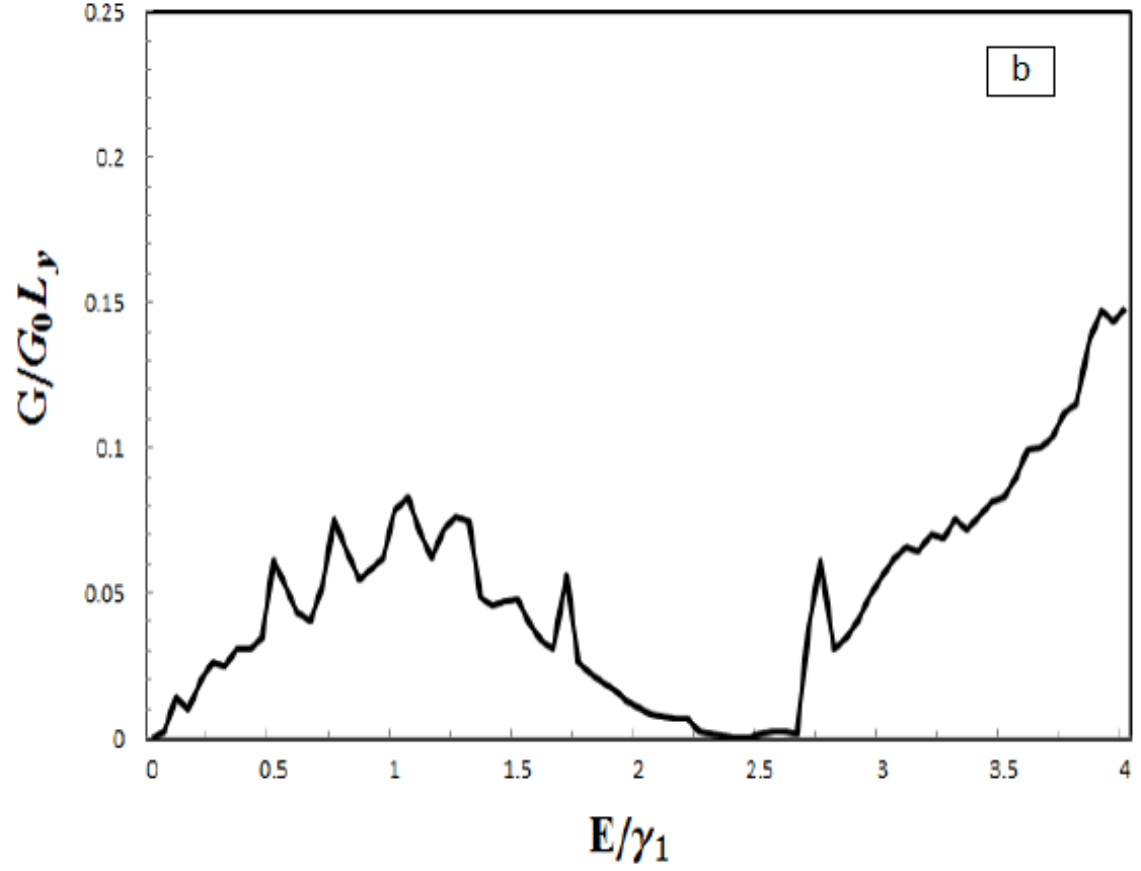}

\caption{Conductance through the single barrier structure in the
presence of a magnetic field as a function of energy for
$V=2.5~\gamma_1$ and $d_2=-d_1=7.5$ nm.
(\textcolor[rgb]{0.98,0.00,0.00}{a}) for $\delta=0.0~\gamma_1$,
$l_B=13.5~nm$ (solid) and $l_B=18.5~nm$ (dotted).
(\textcolor[rgb]{0.98,0.00,0.00}{b}) for $\delta=0.3~\gamma_1$ and
$l_B=13.5~nm$ (solid).}\label{fig.11}
\end{figure}

In Figure \ref{fig.11} we show the conductance through a single
barrier structure in the presence of a magnetic field as a
function of the energy $E$ for $V=2.5~\gamma_1$,
$d_2=-d_1=7.5~nm$ for $l_B=13.5~nm$ (solid) and $l_B=18.5~nm$
(dotted). For energies smaller than the barrier's height, the
peaks in the conductance through a single barrier in the presence of
a magnetic field, which are magnified in the inset of Figure
\ref{fig.11}(\textcolor[rgb]{0.98,0.00,0.00}{a}), have shoulders
due to the presence of resonances in the transmission
probability $T_{+}^+$ in the region $0 < E < V$ and that of
$T_{+}^-$, $T_{-}^+$, and $T_{-}^-$ in the region $\gamma_1 < E <
V$ as depicted in Figure \ref{fig.9}. The resonance peaks of the
conductance resulting from propagation via $\alpha^+$ modes in the
region $E < V-\gamma_1$, appear as shoulders of on other peaks
\cite{Duppen}. Additional resonance peaks appear due to propagation
via $\alpha^-$ modes inside the barrier for energy larger than
$\gamma_1$, $E > \gamma_1$. We should mention the inequality of
the two channels $T_{-}^{+} \neq T_{+}^{-}$ due to the asymmetry
in the presence of the magnetic field. For $V < E < V+\gamma_1$
the contribution of $T_{-}^-$ is zero due to the cloak effect
\cite{Duppen,Hassan}.
To see the effect of the interlayer electrostatic potential, we
plot the conductance as function of the energy $E$ in Figures
\ref{fig.11}(\textcolor[rgb]{0.98,0.00,0.00}{b}) and notice that
the conductance in the energy region $\Delta E = 2 \delta$ is
correlated to the transmission gap.

\section{Conclusion}

In the present work we computed the transmission probability through rectangular
potential barriers and p-n junctions in the presence of both electric and
magnetic static fields in bilayer graphene. The tight binding model that describes our system
leads to the formation of four bands in the associated energy spectrum. The richness of
the energy spectrum allows for two propagation modes whose energy scale is set by the interlayer
coupling $\gamma_1$. For energies higher than the interlayer coupling $\gamma_1$, $E~>~\gamma_1$,
two propagation modes are available for transport, and four possible ways for transmission and
reflection coefficients, while, when the energy is less than $\gamma_1$ the Dirac fermions have
only one mode of propagation available to them. The resulting conductance incorporates these new transport
channels which manifest themselves by the presence of more resonances and larger values of the conductance
at high energies.
The presence of an externally controlled electrostatic potential $\delta$ created an asymmetry between
the on-site energies in the two layers which then resulted in a tunable energy gap between the
conduction and valence energy bands. Hence we studied the effect of the interlayer electrostatic
potential $\delta$ and the various barrier geometry parameters on the transmission probability.

\section*{Acknowledgments}

The generous support provided by the Saudi Center for Theoretical Physics (SCTP)
is highly appreciated by all authors. Bahlouli and Jellal acknowledge partial support
by King Fahd University of petroleum and minerals under the theoretical physics
research group project RG1306-1 and RG1306-2.

\renewcommand{\theequation}{A-\arabic{equation}}
\setcounter{equation}{0}
\section*{Appendix: Wavefunction of our system}

Hamiltonian \eqref{eq1} was used in the Schrodinger equation
$H\psi(x,y)=E\psi(x,y)$ which can then be written as four linear
differential equations of the from
\begin{subequations}
\begin{align}\label{eq4a}
{-i\hbar v_F
\frac{\sqrt{2}}{l_B}a\psi_{B_1}(x,k_y)=(E-V-\delta)\psi_{A_1}(x,k_y)\qquad
\qquad\quad\quad\quad~~}\\
\label{eq4b}{i\hbar v_F
\frac{\sqrt{2}}{l_B}a^{+}\psi_{A_1}(x,k_y)=(E-V-\delta)\psi_{B_1}(x,k_y)-\gamma_1\psi_{A_2}(x,k_y)\quad}\\
\label{eq4c}{\quad\quad-i\hbar v_F
\frac{\sqrt{2}}{l_B}a\psi_{B_2}(x,k_y)=(E-V+\delta)\psi_{A_2}(x,k_y)-\gamma_1\psi_{B_1}(x,k_y)~~~}\\
\label{eq4d}{i\hbar v_F
\frac{\sqrt{2}}{l_B}a^{+}\psi_{A_2}(x,k_y)=(E-V+\delta)\psi_{B_2}(x,k_y)\qquad
\qquad\qquad\quad~}
\end{align}
\end{subequations}
where
$a=\frac{l_B}{\sqrt{2}}\left(\partial_x+k_y+\frac{e}{\hbar}A_y(x)\right)$
and
$a^{+}=\frac{l_B}{\sqrt{2}}\left(-\partial_x+k_y+\frac{e}{\hbar}A_y(x)\right)$
are the annihilation and creation operators. We find the
expression of $\psi_{A_1}(x,k_y)$ in  \eqref{eq4a} and
$\psi_{B_2}(x,k_y)$ in  \eqref{eq4d}, and replace
both $\psi_{A_1}(x,k_y)$ and $\psi_{B_2}(x,k_y)$ in
\eqref{eq4b} and \eqref{eq4c}, respectively. This  gives
\begin{subequations}
\begin{align}\label{eq5a}
 {\left(2\vartheta_{0}^2 a^{+}a-(E-V-\delta)^{2}\right)\psi_{B_1}(x,k_y)=-\gamma_1(E-V-\delta)\psi_{A_2}(x,k_y)~~}\\
 {\left(2\vartheta_{0}^2 aa^{+}-(E-V+\delta)^{2}\right)\psi_{A_2}(x,k_y)=-\gamma_1(E-V+\delta)\psi_{B_1}(x,k_y)~~}
\end{align}
\end{subequations}
where $\vartheta_0=\frac{\hbar v_F}{l_B}$ is the energy scale. Combining  the above  equations we obtain
\begin{equation}\label{eqgnrl}
\left[2\vartheta_{0}^2aa^{+}-(E-V+\delta)^{2}\right]\left[2\vartheta_{0}^2
a^{+}a-(E-V-\delta)^{2}\right]\psi_{B_1}(x,k_y)
=\gamma_{1}^{2}((E-V)^2-\delta^2)\psi_{B_1}(x,k_y)
\end{equation}
Solving the eigenvalue equation we end up with the eigenspinors outside
$(x < d_1, x > d_2)$ and inside $(d_1 < x < d_2)$ the barrier
regions which result in the following two situations:

\subsection*{a) Inside the barrier region}

In region ${\sf II}$ ($d_1 < x < d_2)$, the vector potential $A_y(x)$ is
given by $\frac{\hbar}{e l_{B}^2}x$ which can then expressed in
terms of annihilation and creation operators ($a$ and $a^+$). Using
the envelope function $\psi_{B_1}(x,k_y)\equiv\psi_{B_1}(X)$ that
depend on a combination of the variables,
$X=\frac{x}{l_{B}}+k_yl_B$, we can rewrite $a$ and $a^+$ as
follows $a=\frac{1}{\sqrt{2}}\left(\partial_X+X\right)$ and
$a^{+}=\frac{1}{\sqrt{2}}\left(-\partial_X+X\right)$. Our
differential equation becomes
\begin{equation}\label{eqslt}
\left[-\partial_{X}^2+X^2-1-2\lambda_+\right]\left[-\partial_{X}^2+X^2-1-2\lambda_-\right]\psi_{B_1}(X)=0
\end{equation}
where
\begin{equation}
\lambda_{\pm}=-\frac{1}{2}+\frac{(E-V)^2+\delta^2}{2\vartheta_{0}^2}\pm
\frac{\sqrt{(\vartheta_{0}^2-2(E-V)\delta)^2+\gamma_{1}^2
((E-V)^2-\delta^2)}}{2\vartheta_{0}^2}.
\end{equation}
Therefore, the general solution of  \eqref{eqslt}, can be
written as follows
$\psi_{B_1}(Z)=\psi_{B_1}^{+}(Z)+\psi_{B_1}^{-}(Z)$
with
\begin{equation}
\begin{array}{ll}
{\psi_{B_1}^{+}(Z)=c_+ D[\lambda_+,Z]+c_-D[\lambda_+,-Z]}\\
{\psi_{B_1}^{-}(Z)=d_+ D[\lambda_-,Z]+d_-D[\lambda_-,-Z]}
\end{array}
\end{equation}
{and we have set} $Z=\sqrt{2}X$. Using this result in equation \eqref{eq4a},
gives $\psi_{A_1}(Z)=\psi_{A_1}^{+}(Z)+\psi_{A_1}^{-}(Z)$ with
\begin{equation}
\begin{array}{ll}
{\psi_{A_1}^{+}(Z)=c_+\eta_-\lambda_+
D[\lambda_+-1,Z]+c_-\eta_{-}^*\lambda_+
D[\lambda_+-1,-Z]}\\{\psi_{A_1}^{-}(Z)=d_+\eta_-\lambda_-
D[\lambda_--1,Z]+d_-\eta_{-}^*\lambda_-D[\lambda_--1,-Z]}
\end{array}
\end{equation}
where $\eta_{\pm}=\frac{-i\sqrt{2}\vartheta_0}{E-V\pm\delta}$.
Furthermore using both $\psi_{A_1}(x,k_y)$ and $\psi_{B_2}(x,k_y)$
in  \eqref{eq4b} gives
$\psi_{A_2}(Z)=\psi_{A_2}^{+}(Z)+\psi_{A_2}^{-}(Z)$  {such as}
\begin{equation}
\begin{array}{ll}
{\psi_{A_2}^{+}(Z)=c_+ \zeta^+ D[\lambda_+,Z]+c_- \zeta^+
D[\lambda_+,-Z]}\\{\psi_{A_2}^{-}(Z)=d_+ \zeta^-
D[\lambda_-,Z]+d_- \zeta^- D[\lambda_-,-Z]}
\end{array}
\end{equation}
 {and
$\zeta^\pm=\frac{E-V-\delta}{\gamma_1}-\frac{2\vartheta_{0}^2\lambda_\pm}{\gamma_1(E-V-\delta)}$ is introduced}.
Finally, using $\psi_{A_2}$ in  \eqref{eq4d} gives
$\psi_{B_2}(Z)=\psi_{B_2}^{+}(Z)+\psi_{B_2}^{-}(Z)$ with
\begin{equation}
\begin{array}{ll}
{\psi_{B_2}^{+}(Z)=c_+\eta_{+}^*\zeta^+
D[\lambda_++1,Z]+c_-\eta_{+}\zeta^+
D[\lambda_++1,-Z]}\\{\psi_{B_2}^{-}(Z)=d_+\eta_{+}^*\zeta^-
D[\lambda_-+1,Z]+d_-\eta_{+}\zeta^- D[\lambda_-+1,-Z]}.
\end{array}
\end{equation}

\subsection*{b) Outside the barrier region}
Solving the eigenvalue equation \eqref{eqgnrl} to obtain the
eigenspinor in region ${\sf I}$ ($x < d_1$) and in region ${\sf III}$ ($ x >
d_2$), where potential barrier $V$ and interlayer potential
$\delta$ are equal to zero and the associated vector potential
$A_y(x)$ is constant and equal to $\frac{\hbar}{el_{B}^2}d_{1}$
($\frac{\hbar}{e l_{B}^2}d_{2}$) in region ${\sf I}$ (region ${\sf III}$). We
obtain the general solution in a plane-wave form
$\psi_{B_1}(x,k_y)=\psi_{B_1}^{+}(x,k_y)+\psi_{B_1}^{-}(x,k_y)$
with
\begin{equation}
\begin{array}{ll}
{\psi_{B_1}^{+}(x,k_y)=c_+e^{i\alpha_{1,2}^{+}x}+c_- e^{-i\alpha_{1,2}^{+}x}}\\
{\psi_{B_1}^{-}(x,k_y)=d_+e^{i\alpha_{1,2}^{-}x}+d_-e^{-i\alpha_{1,2}^{-}x}}
\end{array}
\end{equation}
where $\alpha_{1,2}^{\pm}=\sqrt{(E^2\pm E\gamma_1)/(\hbar
v_F)^2-\left(k_y+\frac{d_{1,2}}{l_{B}^2}\right)^2}$ is the parallel wave
vector component in the $x$-direction while indices 1 and 2 represent the two regions ${\sf I}$
and ${\sf III}$, respectively. Using this result in  \eqref{eq4a}
gives
$\psi_{A_1}(x,k_y)=\psi_{A_1}^{+}(x,k_y)+\psi_{A_1}^{-}(x,k_y)$
with
\begin{equation}
\begin{array}{ll}
{\psi_{A_1}^{+}(x,k_y)=c_+f_{1,2}^{++}e^{i\alpha_{1,2}^{+}x}+c_-f_{1,2}^{+-}
e^{-i\alpha_{1,2}^{+}x}}\\{\psi_{A_1}^{-}(x,k_y)=d_+f_{1,2}^{-+}e^{i\alpha_{1,2}^{-}x}+d_-f_{1,2}^{--}e^{-i\alpha_{1,2}^{-}x}}
\end{array}
\end{equation}
where
$f_{1,2}^{\pm\pm}=\left(\pm\alpha_{1,2}^{\pm}-i\left(k_y+\frac{d_{1,2}}{l_{B}^2}\right)\right)\hbar
v_F/E$. Replacing both $\psi_{A_1}(x,k_y)$ and $\psi_{B_2}(x,k_y)$
in equation \eqref{eq4b} gives
$\psi_{A_2}(x,k_y)=\psi_{A_2}^{+}(x,k_y)+\psi_{A_2}^{-}(x,k_y)$
with
\begin{equation}
\begin{array}{ll}
{\psi_{A_2}^{+}(x,k_y)=-c_+e^{i\alpha_{1,2}^{+}x}-c_-e^{-i\alpha_{1,2}^{+}x}}\\
{\psi_{A_2}^{-}(x,k_y)=d_+e^{i\alpha_{1,2}^{-}x}+d_-e^{-i\alpha_{1,2}^{-}x}}.
\end{array}
\end{equation}
Finally, we use $\psi_{A_2}$ in  \eqref{eq4d} gives
$\psi_{B_2}(x,k_y)=\psi_{B_2}^{+}(x,k_y)+\psi_{B_2}^{-}(x,k_y)$
with
\begin{equation}
\begin{array}{ll}
{\psi_{B_2}^{+}(Z)=-c_+g_{1,2}^{++}e^{i\alpha_{1,2}^{+}x}-c_-g_{1,2}^{+-}
e^{-i\alpha_{1,2}^{+}x}}\\{\psi_{B_2}^{-}(Z)=d_+g_{1,2}^{-+}e^{i\alpha_{1,2}^{-}x}+d_-g_{1,2}^{--}e^{-i\alpha_{1,2}^{-}x}}
\end{array}
\end{equation}
where
$g_{1,2}^{\pm\pm}=\left(\pm\alpha_{1,2}^{\pm}+i\left(k_y+\frac{d_{1,2}}{l_{B}^2}\right)\right)\hbar
v_F/E$.

\end{document}